\documentclass[aps,prd,twocolumn,showpacs,superscriptaddress,nofootinbib,groupedaddress]{revtex4-1} 
\usepackage{graphicx}
 \usepackage{epsfig}
\usepackage{tabularx}
\usepackage{booktabs}
\usepackage{amssymb}
\usepackage{amsmath}
\RequirePackage[colorlinks,citecolor=blue,urlcolor=blue,linkcolor=blue]{hyperref}
\usepackage{dcolumn}
\begin{document} 
\preprint{PRD}

\title{The mixing angle as a function of neutrino mass ratio}

\author{S.~Roy}
 \email{meetsubhankar@gmail.com}
\affiliation{Department of Physics, Gauhati University, Guwahati, Assam -781014, India
}%
\author{N.~N.~Singh}
 \email{nimai03@yahoo.com}
\affiliation{Department of Physics, Manipur University, Imphal, Manipur-795003, India
}%
\date{\today}


 
\begin{abstract}
In the quark sector, we experience a correlation between the mixing angles and the mass ratios. A partial realization of the similar tie-up in the neutrino sector helps to constrain the parametrization of masses and mixing, and hints for a predictive framework. We derive five hierarchy dependent textures of neutrino mass matrix with minimum number of parameters ($\leq\,4$), following a model-independent strategy.

\end{abstract}

\pacs{11.30.Hv 14.60.-z 14.60.Pq 14.80.Cp 23.40.Bw}
\keywords{Neutrino masses, Neutrino mixing, Cabibbo angle, $\mu$-$\tau$ symmetry}
\maketitle

The neutrino mass matrix plays the central role in the study of neutrino physics, as it contains the information of both the masses and mixing. In that sense, it is more fundamental than the PMNS matrix. It is always desired to derive a texture of the mass matrix which leads to significant prediction. If the neutrino mass matrix, $\mathcal{M_{\nu}}$ follows $\mu$-$\tau$ symmetry\, \citep{Fukuyama:1997ky, Mohapatra:1998ka, Lam:2001fb, Mohapatra:2006pu, Harrison:2002et, Grimus:2001ex, Grimus:2003kq, Kitabayashi:2002jd, Koide:2003rx} , we obtain two constraints on the matrix elements; they are: $(\mathcal{M_{\nu}})_{12}=(\mathcal{M_{\nu}})_{13}$ and   $(\mathcal{M_{\nu}})_{22}=(\mathcal{M_{\nu}})_{33}$. These two constraints generate : $\theta_{13}=0$ and $\theta_{23}=45^0$. But the $\mu$-$\tau$ symmetric texture does not tell anything about the neutrino mass hierarchy and solar angle. The texture becomes predictive only when it is associated with certain flavor symmetries\,\citep{Ma:2001dn,Ma:2002yp,Ma:2004zv,Ma:2005qf,Altarelli:2005yp,Altarelli:2005yx,Grimus:2009pg,
Zee:2005ut,Adhikary:2006jx}. On the contrary, the present experimental data strongly rule out any possibility of a vanishing reactor angle\,\citep{Ahn:2012nd,Abe:2011fz,An:2012eh} and the central value of $\theta_{23}$ is more than $45^0$ (and is close to $49^0$)\,\citep{Forero:2014bxa}. These deviations undoubtedly questions the credibility of $\mu$-$\tau$ symmetry.  

Visualization of a more realistic neutrino mixing pattern and mass matrix, demands perturbation to the $\mu$-$\tau$ symmetry\,\citep{Adhikary:2012mt, Grimus:2012hu,Adhikary:2013mfa}. In the present article we emphasize more on the possibility to perceive an exact texture of $M_{\nu}$ which is \textit{model independent}, with minimum number of efficient elements, than following perturbation techniques. Our approach is bottom-up and inspired by the phenomenology of quark sector.

The Cabibbo angle ($\theta_{c}$)\,\citep{Cabibbo:1977nk, Wolfenstein:1983yz} is a parameter which plays a significant role in describing the quark masses and mixing. It is anticipated that this angle might be a function of the ratio of down and strange quark masses\,\citep{Gatto:1968ss},

\begin{eqnarray}
\sin\theta_{c}\simeq\sqrt{\frac{m_{d}}{m_{s}}}.
\end{eqnarray} 

It is an esteemed endeavor of particle physicists to unify the quark and lepton sectors, or to realize similar kind of happenings in both the sectors otherwise. Based on this, is it possible to extend a similar idea in the form of an \textit{ansatz} in the neutrino sector also, as in the following,
\begin{eqnarray}
\label{ansatz}
\sqrt{\frac{m_i}{m_j}}=\sin\theta_{ij}.\quad (i,j=1,2,3)\,?
\end{eqnarray}

Undoubtedly there are several hurdles which will arise both from the theoretical and phenomenological perspectives. The reason being the difference between the mixing mechanism in both the sectors. The CKM matrix is very close to unit matrix and the spectra of ``up'' and ``down'' quarks are strongly hierarchical. But, for neutrinos we are ignorant of the exact hierarchy of the masses. Unlike the quarks, the mixing is quite large in lepton sector and the PMNS matrix is far from being an unit matrix. 

\begin{figure*}
\center{
\includegraphics[scale=0.53]{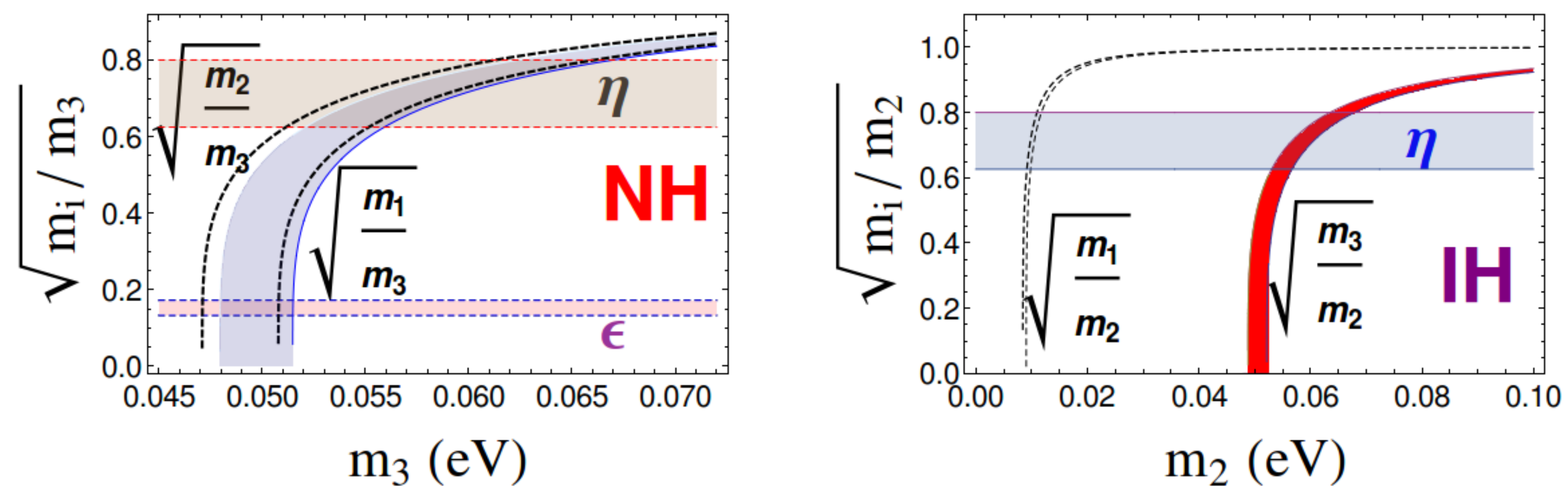}
\caption{\label{fig1}\footnotesize{The evolution of the neutrino mass ratios with respect to the absolute mass scale for both normal (\textbf{left})and inverted ordering (\textbf{right}) of the neutrino masses (Corresponding to $3\sigma$ range of $\Delta m_{21}^2$ and $\Delta m_{31}^2$) are illustrated.}} }
\end{figure*}

So long the reactor angle was predicted to be vanishing, such development in Eq.(\ref{ansatz}) seems obsolete. But, in the light of present data, when, $\theta_{13}\sim\theta_{c}$, one cannot deny the possible existence of the following relation,

\begin{eqnarray}
\label{13}
\sqrt{\frac{m_1}{m_3}}\simeq\sin\theta_{13}=\epsilon,  \quad \text{(say)},
\end{eqnarray} 
in the non-degenerate spectrum of neutrino masses, obeying normal ordering (see Fig.(\ref{fig1})). 
We emphasize that the realization of the ansatz in Eq.\,(\ref{ansatz}) is not ``full'', but ``\textit{partial}''. Because, it seems impossible to realize all the three possibilities (See in Eq.\,(\ref{ansatz})) \textit{simultaneously}. For example, if we realize Eq.\,(\ref{13}), then another possibility,
\begin{eqnarray}
\label{23}
\sqrt{\frac{m_2}{m_3}}\simeq\sin\theta_{23}=\eta;
\end{eqnarray} 
is ruled out and vice-versa (See Fig.\,(\ref{fig1})). Again, a similar realization in $1$-$2$ sector is forbidden because of the smallness of the solar mass squared difference, which insists the mass ratio, $m_{1}:m_{2}$ to be constant, and this ratio tends towards unity.

Because of similar reasons, for inverted ordering of the neutrino masses, only the $2$-$3$ realization, 

\begin{eqnarray}
\label{32}
\sqrt{\frac{m_3}{m_2}}\simeq\sin\theta_{23}=\eta;
\end{eqnarray} 
is possible.

Now even though there are several clampdowns, a partial realization of the ansatz (Eq.\,(\ref{ansatz})) clutches certain positive aspects like it puts some constraints on the parametrization of neutrino masses and mixing. Let us discuss some implications of the ansatz in Eq.\,(\ref{ansatz}).

\begin{itemize}

\item It categorizes the parametrization in both ways: ``$\epsilon$-based'' or ``$\eta$-based''.
\item The ``$\epsilon$-based'' parametrization only encompasses the Normal ordering of the masses with non-degenerate spectrum (\textbf{NH-ND}), with absolute mass scale, $ 0.047\,eV\leq m_{3}\leq 0.05\,eV$. This parametrization rules out any possibility of vanishing $m_{1}$ (Strict-NH case). Since the reactor angle is not zero  as depicted in Eq.\,(\ref{13}).
\item ``$\eta$-based'' parametrization encompasses both the  Normal ordering and inverted ordering of neutrino masses with degenerate spectrum (\textbf{NH-QD} and \textbf{IH-QD}) only, with absolute mass scale,  $ 0.05\,eV\leq m_{3} (\,m_{2}\,)\leq 0.067\,eV$.
\item The present ansatz (Eq.\,(\ref{ansatz})), rules out the possibility of non-degenerate inverted spectrum of neutrino masses.
\item In the degenerate limit, the ansatz sets the upper limit on the sum of the neutrino masses,  $\Sigma\, m_{i}\leq 0.17\,eV$. This prediction is relevant in the light of present Cosmological observation. 
\end{itemize}

We can see that although the ansatz, (Eq.\,(\ref{ansatz})) cannot solve the hierarchy issue, yet it can put some constraints on the mass spectrum. This are subjected to the sensitivities of the future experiments. The $\epsilon$ and $\eta$ based mass spectrum can be represented in the following way,

\begin{eqnarray}
M_{\nu}^{d}(\epsilon,s)&=&\begin{bmatrix}
s \epsilon^2 & 0          & 0\\
  0        & s \epsilon & 0\\
  0        & 0          & 1
\end{bmatrix}m_3, \hspace{0.7cm} \text{\textbf{(NH-ND)}}\\
M_{\nu}^{d}(\eta,c)&=&\begin{bmatrix}
c \eta^2 & 0          & 0\\
  0        &  \eta^2 & 0\\
  0        & 0          & 1
\end{bmatrix}m_3, \hspace{0.7 cm} \text{\textbf{(NH-QD)}}\\
M_{\nu}^{d}(\eta,c)&=&\begin{bmatrix}
c  & 0          & 0\\
  0        & 1 & 0\\
  0        & 0          & \eta^2
\end{bmatrix}m_2, \hspace{1 cm} \text{\textbf{(IH-QD)}}
\end{eqnarray} 
where, $s$ and $c$ are $\mathcal{O}(1)$ coefficients: $\epsilon< s< \epsilon^{-1}$, $\eta< c< \eta^{-1}$. The positivity of $\Delta\,m_{21}^2$ enforces, $c\,<\,1$.

We are working in a basis, where the charged lepton mass matrix is diagonal. We represent the PMNS matrix, $U(\epsilon)$ which is ``$\epsilon$'' motivated as in the following,
\begin{widetext}
\begin{eqnarray}
\label{U1}
U(\epsilon,d,f)\approx
 \begin{bmatrix}
 1-\frac{1}{2} f^2 \epsilon ^2-\frac{\epsilon ^2}{2} & f \epsilon  & \epsilon  \\
-f \epsilon  -d \epsilon ^2 & 1-\frac{1}{2} f^2 \epsilon ^2-\frac{d^2 \epsilon ^2}{2} & d \epsilon  \\
 -\epsilon+c d \epsilon ^2  & -d \epsilon-f \epsilon ^2  & 1 -\frac{1}{2} d^2 \epsilon ^2-\frac{\epsilon ^2}{2} \\
\end{bmatrix}, \hspace{1 cm}\text{(\textbf{PMNS-I})}
\end{eqnarray}
\end{widetext}
Where, $d$ and $f$ are $\mathcal{O}(1)$ coefficients. We put forward another possible form of PMNS matrix, which is motivated by $\eta$-based parametrization, $U(\eta)$,

\begin{widetext}
\begin{eqnarray}
\label{U2}
U(\eta,b,c)\approx 
\begin{bmatrix}
 \sqrt{\frac{2}{3}} c' & \frac{c}{\sqrt{3}} & b \gamma  \\
 -\frac{c \kappa }{\sqrt{3}}-\sqrt{\frac{2}{3}} b \gamma  \eta  c' & \sqrt{\frac{2}{3}} \kappa  c'-\frac{b c \gamma  \eta }{\sqrt{3}} & \eta  \\
 \frac{c \eta }{\sqrt{3}}-\sqrt{\frac{2}{3}} b \gamma  \kappa  c' & -\frac{b c \gamma  \kappa }{\sqrt{3}}-\sqrt{\frac{2}{3}} \eta  c' & \kappa  \\
\end{bmatrix}\hspace{1.3 cm}\text{(\textbf{PMNS-II})}
\end{eqnarray}
\end{widetext}

where, $\gamma(\eta)=\eta^8$, $\kappa (\eta)=\cos\sin^{-1}(\eta)$ and $c'=(3-c^2)^{\frac{1}{2}}/2$. 
Let us summarize some important features of the above two non familiar parametrization of PMNS matrix.
\begin{itemize}
\item This is to be highlighted that in either of the two possibilities \textbf{PMNS-I} or \textbf{PMNS II}, a vanishing reactor angle is not possible. If this is so, the mass eigenvalues will also disappear. Hence, the present parametrization cannot hold the Tri-Bimaximal (TB) and Bi-maximal (BM) framework in exact form, which assumes predominantly the reactor angle to be zero. 

\item The \textbf{PMNS-II} allows only the possibilities, $\theta_{23}> 45^0$ and $\theta_{12}\lesssim \sin^{-1}(1/\sqrt{3})$.

\item The free parameter, $\epsilon\sim\mathcal{O}(\lambda)$ and $\eta\sim 4\,\lambda$, where $\lambda$ is the Wolfenstein parameter.
\end{itemize} 

Next, we try to understand how the ansatz in Eq.\,(\ref{ansatz}) will help to understand the texture of neutrino mass matrix? The ansatz reduces the number of free parameters, and the number of working parameters are less than that of physical ones. Hence we expect that the mass matrix is little predictive. To construct the same, we concentrate on the finding out some \textit{exact} sum rules  to relate different matrix elements. 

Here, we construct the neutrino mass matrix, $\mathcal{M_{\nu}=U.M_{\nu}}^{d}.U^T$. For a lucid flow of the present discussion we keep aside the numerical description of the internal parameters like $\epsilon$, $\eta$ etc. The details of the same can be found in the Table.(\ref{table4}). We have the general texture of left-handed Majorana neutrino mass matrix, as shown below.
 
\begin{eqnarray}
M_{\nu}\sim\begin{bmatrix}
\mathcal{A} & \mathcal{B} & \mathcal{C} \\
\mathcal{B}& \mathcal{D}  &  \mathcal{E}\\
 \mathcal{C} & \mathcal{E}& \mathcal{F}
 \end{bmatrix}.
\end{eqnarray}

If, the parametrization is $\epsilon$-based, we have $\mathcal{M_{\nu}}=\mathcal{M_{\nu}}(\epsilon,d,f,s)$ and $\mathcal{M_{\nu}}=\mathcal{M_{\nu}}(\eta,b,c)$, if it is $\eta$-based. We start with the \textbf{NH-ND} case, parametrization of which is $\epsilon$ based. The matrix elements are tabulated in Table.\,(\ref{table1}).
\begin{table*}
\begin{center}
\begin{tabular}{l}
\hline
\hline
\vspace{0.1 cm}
{\footnotesize $\mathcal{A}\approx -f^2 s \epsilon ^4+f^2 s \epsilon ^3-s \epsilon ^4+s \epsilon ^2+\epsilon ^2$ }\\ 
\vspace{0.1 cm}
{\footnotesize $\mathcal{B}\approx -\frac{1}{2} d^2 f s \epsilon ^4-d s \epsilon ^4-\frac{d \epsilon ^4}{2}+d \epsilon ^2-\frac{1}{2} f^3 s \epsilon ^4-\frac{1}{2} f s \epsilon ^4-f s \epsilon ^3+f s \epsilon ^2$ }\\ 
\vspace{0.1 cm}
{\footnotesize $\mathcal{C}\approx -\frac{d^2 \epsilon ^3}{2}+d f s \epsilon ^4-d f s \epsilon ^3-f^2 s \epsilon ^4-s \epsilon ^3-\frac{\epsilon ^3}{2}+\epsilon $ }\\ 
\vspace{0.1 cm} 
{\footnotesize $\mathcal{D}\approx -d^2 s \epsilon ^3-d^2 \epsilon ^4+d^2 \epsilon ^2-2 d f s \epsilon ^4+f^2 s \epsilon ^4-f^2 s \epsilon ^3+s \epsilon$ }\\ 
\vspace{0.1 cm}
{\footnotesize $\mathcal{E}\approx \frac{1}{2} d^3 s \epsilon ^4-\frac{d^3 \epsilon ^3}{2}+d f^2 s \epsilon ^4-d s \epsilon ^2-d \epsilon ^3+d \epsilon +f s \epsilon ^4-f s \epsilon ^3$} \\ 
\vspace{0.1 cm}
{\footnotesize $\mathcal{F}\approx d^2 s \epsilon ^3+d^2 \epsilon ^4-d^2 \epsilon ^2+2 d f s \epsilon ^4+s \epsilon ^4+\frac{\epsilon ^4}{4}-\epsilon ^2+1$} \\ 
\hline 
\end{tabular} 
\caption{\label{table1} The elements of the general neutrino mass matrix (Normal Hierarchy-non-degenerate (\textbf{NH-ND}) case)}
\end{center}
\end{table*}

Choosing the working parameters, ($\epsilon$, $d$, $f$, $s$) properly, we derive the following exact relations binding the matrix elements.

\begin{eqnarray}
&\text{Sum rule 1:}&\quad 2 (\mathcal{A}+\mathcal{C})-(\mathcal{E}-\mathcal{B})=0,\\
&\text{Sum rule 2:}&\quad \mathcal{D}-(\mathcal{A}+\mathcal{F})=0,\\
&\text{Sum rule 3:}&\quad 2 \mathcal{A}-(\mathcal{B}+\mathcal{C})=0,\\
&\text{Sum rule 4:}&\quad 2 \mathcal{A}-(\mathcal{D}-\mathcal{E})=0,
\end{eqnarray}

These sum-rules promote a framework with,
\begin{eqnarray}
\theta_{13}\approx 8.73^0,\quad\theta_{23}\approx 48.96^0, \quad \theta_{12}\approx 32.51^0.
\end{eqnarray}
Except the solar angle which is consistent with $2\sigma$ range (by sacrificing one of the sum rules, the solar angle can be made precise.), the rest lies within $1\sigma$ range. The above sum rules lead to the following texture,

\begin{eqnarray}
M_{\nu}&=&\begin {bmatrix}
\mathcal {A} & 2 \,\mathcal {A} - \mathcal {C} & \mathcal {C} \\ 
2\,\mathcal {A} - \mathcal {C} & 6\,\mathcal {A} + \mathcal {C} & 4\, \
\mathcal {A} + \mathcal {C} \\ 
\mathcal {C} & 4\, \mathcal {A} + \
\mathcal {C} & 5\, \mathcal {A} + \mathcal {C} \\
\end {bmatrix}m_{3},\,\, \text{(\textbf{Texture-I})}\nonumber\\
\end{eqnarray} 

where, $\mathcal{A}>\mathcal{C}$. It is interesting to note that the \textbf{Texture-I} is nothing but combination of a pattern akin to the modified Fritzsch-like texture of quark mass matrices\,\citep{Fritzsch:1999yd,Fritzsch:1999ee},

\begin{eqnarray}
\begin {bmatrix}
0 & 2 \,\mathcal {A}  & 0 \\ 
2\,\mathcal {A}  & 6\,\mathcal {A} & 4\, \mathcal {A}  \\ 
0 & 4\, \mathcal {A}  & 5\, \mathcal {A}  \\
\end {bmatrix},
\end{eqnarray}

and, a $\mu$-$\tau$ symmetric texture as shown below,

\begin{eqnarray}
\begin {bmatrix}
\mathcal {A} &  - \mathcal {C} & \mathcal {C} \\
- \mathcal {C} &  \mathcal {C} & \mathcal {C} \\ 
\mathcal {C} & \mathcal {C} &  \mathcal {C}\\
\end {bmatrix}.
\end{eqnarray}

Next, we turn towards another possibility, i.e., the \textbf{NHQD} scenario which is motivated by $\eta$-based parametrization. The matrix elements of the concerned neutrino mass matrix are illustrated in Table.(\ref{table2}).

\begin{table*}
\begin{center}
\begin{tabular}{l}
\hline
\hline
\vspace{0.1 cm}
{\footnotesize $\mathcal{A}\approx b^2 \gamma ^2+\frac{c^2 \eta ^2}{3}+\frac{2}{3} c \eta ^2 c'^2$ }\\ 
\vspace{0.1 cm}
{\footnotesize $\mathcal{B}\approx -\frac{1}{3} \eta  \{b \gamma  \left(c^2 \eta ^2-3\right)+2 b c \gamma  \eta ^2 c'^2+\sqrt{2} (c-1) c \eta  \kappa  c'\}$ }\\ 
\vspace{0.1 cm}
{\footnotesize $\mathcal{C}\approx \frac{1}{3} \{b \gamma  \kappa  \left(3-c^2 \eta ^2\right)-2 b c \gamma  \eta ^2 \kappa c'^2+\sqrt{2} (c-1) c \eta ^3 c'\}$ }\\ 
\vspace{0.1 cm} 
{\footnotesize $\mathcal{D}\approx \frac{1}{9} \eta ^2 \{c \left(\sqrt{6} b \gamma  \eta  c'+\sqrt{3} c \kappa \right)^2+3 \left(b c \gamma  \eta -\sqrt{2} \kappa  c'\right)^2+9\}$ }\\ 
\vspace{0.1 cm}
{\footnotesize $\mathcal{E}\approx-\frac{1}{3} \eta  \{-2 \eta ^2 \kappa  \left(c'\right)^2 \left(b^2 c \gamma ^2-1\right)+\kappa  \left(-b^2 \gamma ^2 c^2 \eta ^2+c^3 \eta ^2-3\right)+\sqrt{2} b (c-1) c \gamma  \eta  c' \left(\eta ^2-\kappa ^2\right)\}$} \\ 
\vspace{0.1 cm}
{\footnotesize $\mathcal{F}\approx \frac{1}{9} \eta ^2 \left(\sqrt{3} b c \gamma  \kappa +\sqrt{6} \eta  c'\right)^2+\frac{1}{9} c \eta ^2 \left(\sqrt{3} c \eta -\sqrt{6} b \gamma  \kappa  c'\right)^2+\kappa ^2$} \\ 
\hline 
\end{tabular} 
\caption{\label{table2} The elements of the general neutrino mass matrix (\textbf{NH-QD} case) }
\end{center}
\end{table*}

As before, we encounter the following exact sum rules,
\begin{eqnarray}
&\text{Sum rule 1:}&\quad \mathcal{D}-\mathcal{F}-\mathcal{B}=0,\\
\label{sum2}
&\text{Sum rule 2:}&\quad 4\, \mathcal{E}-\mathcal{D}=0,\\
&\text{Sum rule 3:}&\quad 3\,\mathcal{C}-2(\mathcal{D}-\mathcal{F})=0,
\end{eqnarray}

which prescribe the following texture of the neutrino mass matrix guided by three parameters,

\begin{eqnarray}
M_{\nu}=\begin {bmatrix}
\mathcal {A} & \mathcal {B} &\frac{2}{3}\mathcal{B}\\
 \mathcal {B} & 4 \mathcal {E} & \mathcal {E} \\
\frac{2}{3}\mathcal{B} & \mathcal {E} & 4 \mathcal {E} - \mathcal {B} \\
\end{bmatrix}m_{3}, \hspace{0.4 cm}\text{(\textbf{Texture-II})}
\end{eqnarray}
with, $4\,\mathcal{E}>\mathcal{A}>\mathcal{E}>\mathcal{B}$. The above neutrino mass matrix is consistent with the prediction,
\begin{eqnarray}
\theta_{12}\approx 34.08^0,\quad \theta_{23}\approx 49.66^0, \quad \theta_{13}\approx 10^0.
\end{eqnarray}

We see that the above framework predicts a reactor angle, lying slightly above the experimental observation. Here also, we experience a $\mu$-$\tau$ symmetric texture with a perturbation matrix,

\begin{eqnarray}
\begin {bmatrix}
\mathcal {A} & \mathcal {B} &\mathcal{B}\\
 \mathcal {B} & 4 \mathcal {E} & \mathcal {E} \\
\mathcal{B} & \mathcal {E} & 4 \mathcal {E} \\
\end{bmatrix} + \begin {bmatrix}
0 & 0 &\frac{1}{3}\mathcal{B}\\
0 & 0 & 0\\
\frac{1}{3}\mathcal{B} & 0 & - \mathcal {B} \\
\end{bmatrix}.
\end{eqnarray}

\begin{figure*}
\center{
\includegraphics[scale=0.5]{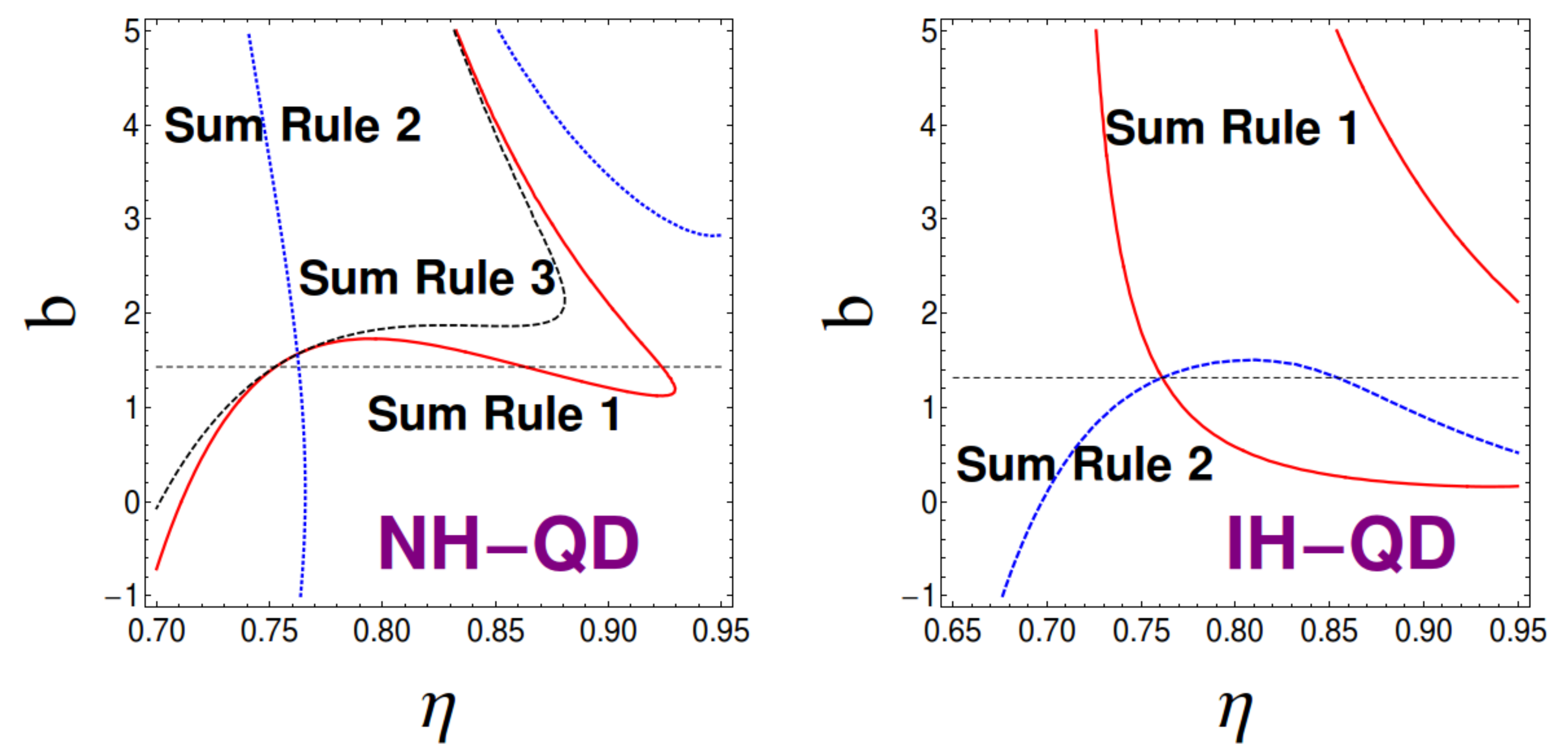}
\caption{\label{fig2}\footnotesize{The illustration of the Sum rules, in the plane $b$-$\eta$  for the mass matrices corresponding to \textbf{Texture-II} (\textbf{left})  and  \textbf{Texture-V} (\textbf{right}). }} }
\end{figure*}

 A precise value $\theta_{13}\approx 9.3^0$, consistent within $1\sigma$ bound is obtainable at the cost of sacrificing the Sum rule 2 in Eq.\,(\ref{sum2}) (See Fig.(\ref{fig2})). The other angles remain untouched. And, the corresponding texture of the neutrino mass matrix appears as in the following,

\begin{eqnarray}
M_{\nu}=\begin {bmatrix} 
\mathcal {A} & \mathcal {B} &\frac {2 } {3} \mathcal{B} \\ 
\mathcal {B} & \mathcal {D} &\mathcal {E} \\
\frac {2 }{3}\mathcal {B} &\mathcal {E} & \mathcal {D} - \mathcal {B} \\
\end{bmatrix}m_{3}, \hspace{0.4 cm}\text{(\textbf{Texture-III})}
\end{eqnarray}

with $\mathcal{D}>\mathcal{A}>\mathcal{E}>\mathcal{B}$. The discussion related to the hidden $\mu$-$\tau$ symmetric texture is similar to that for \textbf{Texture-II}.

Similarly for the case of Inverted hierarchy (\textbf{IH-QD}), the matrix elements are shown in Table.(\ref{table3}). The realization of the following sum rules:

\begin{eqnarray}
&\text{Sum rule 1:}&\quad \mathcal{B}-\mathcal{C}=0,\\
&\text{Sum rule 2:}&\quad \mathcal{A}+\mathcal{E}-\mathcal{D}+\mathcal{B}=0,
\end{eqnarray}

promote the neutrino mass matrix, $M_{\nu}$, to assume the following texture,

\begin{eqnarray}
M_{\nu}=\begin {bmatrix} 
\mathcal {A} &\mathcal {B} & \mathcal {B} \\ 
\mathcal {B} &\mathcal { 
      D} &\mathcal {D} - \mathcal {A} - \mathcal {B} \\ 
B &\mathcal {D} - \mathcal {A} - \mathcal {B} &  \mathcal {F} \\
\end {bmatrix}m_{2},\text{(\textbf{Texture-IV})}\nonumber\\
\end{eqnarray}

with $\mathcal{A}> \mathcal{F}>\mathcal{D}>|\mathcal{B}|$.

\begin{table*}
\begin{center}
\begin{tabular}{l}
\hline
\hline
\vspace{0.1 cm}
{\footnotesize $\mathcal{A}\approx b^2 \gamma ^2 \eta ^2+\frac{c^2}{3}+\frac{2}{3} c c'^2$ }\\ 
\vspace{0.1 cm}
{\footnotesize $\mathcal{B}\approx -\frac{1}{3} b c^2 \gamma  \eta -\frac{2}{3} b c \gamma  \eta  c'^2+b \gamma  \eta ^3+\frac{1}{3} \sqrt{2} c \kappa  c'-\frac{1}{3} \sqrt{2} c^2 \kappa  c'$ }\\ 
\vspace{0.1 cm}
{\footnotesize $\mathcal{C}\approx -\frac{1}{3} b \gamma  c^2 \kappa -\frac{2}{3} b \gamma  c \kappa c'^2+b \gamma  \eta ^2 \kappa -\frac{1}{3} \sqrt{2} c \eta  c'+\frac{1}{3} \sqrt{2} c^2 \eta  c'$ }\\ 
\vspace{0.1 cm} 
{\footnotesize $\mathcal{D}\approx \frac{1}{3} b^2 c^2 \gamma ^2 \eta ^2+\frac{2}{3} b^2 c \gamma ^2 \eta ^2 c'^2-\frac{2}{3} \sqrt{2} b c \gamma  \eta  \kappa  c'+\frac{1}{3} \sqrt{2} 2 b c^2 \gamma  \eta  \kappa  c'+\frac{c^3 \kappa ^2}{3}+\frac{2}{3} \kappa ^2 c'^2+\eta ^4$ }\\ 
\vspace{0.1 cm}
{\footnotesize $\mathcal{E}\approx\frac{1}{3} b^2 \gamma ^2 c^2 \eta  \kappa +\frac{2}{3} b^2 \gamma ^2 c \eta  \kappa c'^2+\frac{1}{3} \sqrt{2} b \gamma  c \eta ^2 c'-\frac{1}{3} \sqrt{2} b \gamma  c \kappa ^2 c'-\frac{1}{3} \sqrt{2} b \gamma  c^2 \eta ^2 c'+\frac{1}{3} \sqrt{2} b \gamma  c^2 \kappa ^2 c'-\frac{1}{3} c^3 \eta  \kappa -\frac{2}{3} \eta  \kappa  c'^2+\eta ^3 \kappa$} \\ 
\vspace{0.1 cm}
{\footnotesize $\mathcal{F}\approx \frac{1}{3} b^2 \gamma ^2 c^2 \kappa ^2+\frac{2}{3} b^2 \gamma ^2 c \kappa ^2 c'^2+\frac{1}{3} \sqrt{2} 2 b \gamma  c \eta  \kappa  c'-\frac{2}{3} \sqrt{2} b \gamma  c^2 \eta  \kappa  c'+\frac{c^3 \eta ^2}{3}+\frac{2}{3} \eta ^2 c'^2+\eta ^2 \kappa ^2$} \\ 
\hline 
\end{tabular} 
\caption{\label{table3} The elements of the general neutrino mass matrix (\textbf{IH-QD} case)}
\end{center}
\end{table*}

The above sum rules, restricts the reactor angle at,
\begin{eqnarray}
\theta_{13}\approx 7.1^0,
\end{eqnarray}
Which is little lower than what we observe experimentally.The other predictions are,
\begin{eqnarray}
\theta_{12}\approx 34.84^{0},\quad \theta_{23}\approx 50.76^0.
\end{eqnarray}
which are consistent within $1\sigma$. By changing the Sum rules a little, (See Fig.(\ref{fig2})) as in the following,
\begin{eqnarray}
&\text{Sum rule 1:}&\quad \mathcal{B}-\mathcal{C}=0,\\
&\text{Sum rule 2:}&\quad \mathcal{A}-\mathcal{B}-\mathcal{F}+\mathcal{E}=0.
\end{eqnarray}
We can set the reactor angle within the $1\sigma$ bound ($\theta_{13}\approx 8.5^0$). The corresponding neutrino mass matrix appears as in the following,

\begin{eqnarray}
M_{\nu}=\begin {bmatrix} 
\mathcal {A} &\mathcal {B} & \mathcal {B} \\ 
\mathcal {B} &\mathcal { 
      D} &\mathcal {F} - \mathcal {A} + \mathcal {B} \\ 
B &\mathcal {F} - \mathcal {A} + \mathcal {B} &  \mathcal {F} \\
\end {bmatrix}m_{2}, \text{(\textbf{Texture-V})}\nonumber\\
\end{eqnarray}

Where, $\mathcal{A}>\mathcal{F}>\mathcal{D}>|\mathcal{B}|$. We know that the TB mixing which is very often studied under several symmetry groups (e.g $A_{4}$, $S_{4}$ and $\Delta(54)$ etc.), compels the elements of neutrino mass matrix to obey three sum rules \citep{Harrison:2002er, Harrison:2004uh, Abbas:2010jw} as shown below,
\begin{eqnarray}
\label{S1}
(M_{\nu})_{12}&=&(M_{\nu})_{13},\\
\label{S2}
(M_{\nu})_{22}&=&(M_{\nu})_{23},\\
\label{S3}
(M_{\nu})_{11}+(M_{\nu})_{12}&=&(M_{\nu})_{22}+(M_{\nu})_{23}.
\end{eqnarray}

One sees, that both \textbf{Texture IV} and \textbf{V} respects the first sum rule (See Eq.\,(\ref{S1})), but breaks the second one (Eq.\,(\ref{S2})). The sum rule in Eq.\,(\ref{S3}) is modified to,
\begin{eqnarray}
(M_{\nu})_{11}-(M_{\nu})_{12}&=&(M_{\nu})_{22}-(M_{\nu})_{23},\\
(M_{\nu})_{12}-(M_{\nu})_{11}&=&(M_{\nu})_{23}-(M_{\nu})_{33},
\end{eqnarray} 
for \textbf{Texture-IV} and \textbf{V} respectively.

In the above discussion, for all the five textures, the parameters are chosen in such a way that the two observational mass parameters can lie always within the $1\sigma$ boundary of experimental data. Also, it is seen that the  number of independent matrix elements are $2$, $3$ or $4$. Needless to say, that the textures displayed above are hierarchy dependent and exact. Also, the interesting fact that one sees is an unavoidable presence of a $\mu$-$\tau$ symmetric or partially broken $\mu$-$\tau$ symmetric textures in the above patterns. But this does not allow us to take the initial choice of $M_{\nu}$ as $\mu$-$\tau$ symmetric, because both $\epsilon$-based or $\eta$-based parametrization are reluctant to assume a vanishing $\theta_{13}$ (which is one of the important traces of the $\mu$-$\tau$ symmetry). Certainly, some other efficient possibilities are to be contrived. 

Although the five possible forms of neutrino mass matrix materialized above are model-independent, yet this is a natural quest whether the sum rules or the associated textures are motivated in flavor symmetries or not. To unveil the first principle working behind these sum rules, is beyond the scope of present note. Alternatively, we try to put forward a noble way to feel the essence of possible symmetries by expressing individual mass matrices explicitly in terms of certain building block textures (see Table.(\ref{table5})). Instead of describing how these building blocks appear we shall describe the physical situations where they may appear (this may help the model-builders to find out certain linkage). We see,

\begin{itemize}
\item The \textbf{Texture-I} is a linear superposition of $H$, $f_{1}$, $d_{1}$, $e_{4}$ and $T_{1}$. The matrix $H$ resembles a ``Fritzsch-like'' texture\citep{Fritzsch:1999yd, Fritzsch:1999ee, Fritzsch:2012zp} which in general appears in the context of quarks. But, the same Fritzsch-like texture is obtainable for neutrinos also using $S_{3}$ symmetry\citep{Meloni:2010aw}. The other matrices, $f_{1}$, $d_{1}$, $e_{4}$ which span the $\mu$-$\tau$ symmetric part of \textbf{Texture-I}, are the members of $S_{4}$ symmetry group and belong to the same conjugacy class $C_{6}$ \citep{King:2014nza}. A partial realization of this $\mu$-$\tau$ symmetric texture is experienced in some works motivated by $S_{4}$\citep{Ishimori:2010fs} and $A_{4}$\citep{Shimizu:2011xg} symmetry groups.

\item The matrix $d_{1}$ again appears in \textbf{Textures II} and \textbf{III} and is followed by the identity matrix, $I$ and the democratic matrix, $D$ (derivable from $S_{3}$ symmetry). A linear combination of $d_1,\,\, D$ and $I$ of such kind, in general contributes towards the mass matrix following TBM mixing and is also realized in different discrete symmetry based models. A framework with $A_{4}\times Z_{3}$ symmetry\citep{Altarelli:2005yp, Altarelli:2005yx} is one such example.

\item The matrix $D$ and $I$ also appear in \textbf{Texture-V}, supplemented by a matrix $d_{2}$, which belongs to $S_{4}$ symmetry group\citep{King:2014nza}. On the other hand, in \textbf{Texture-IV}, $D$ and $I$ are replaced by $c_{3}$ and $b_{3}$, which are the members of $A_{4}$ symmetry\citep{King:2014nza}. 

\item The matrix $T_{1}$, appears in \textbf{Texture-I, II, III} and its presence is experienced in the neutrino mass matrix, where a deviation from TBM scenario \citep{Abbas:2010jw} is taken into consideration.  The matrix $T_{2}$, plays an important role in the $\mu$-$\tau$ symmetry violation for \textbf{Texture-II} and \textbf{III}. In Refs.\citep{Adhikary:2012mt,Adhikary:2012zx}, one sees that to break the $\mu$-$\tau$ symmetry of a neutrino mass matrix (concerning texture zeros), a matrix of following kind,
\begin{eqnarray}
\begin{bmatrix}
-k_{2}^2 & 0 & -k_{2}\\
0 & 0 & 0\\
-k_{2} & 0& -1 
\end{bmatrix},
\end{eqnarray} 
emerges and it may resemble $T_{2}$, if we assume $k_{2}=1/3$.
\item The matrices $T_{3}$ and $T_{4}$, present in \textbf{Texture-IV} and \textbf{V} respectively, do not obey $\mu$-$\tau$ symmetry. The parameters $\mathcal{F}$ and $\mathcal{D}$ are comparable and may obey the same in the limit, $\mathcal{F}/\mathcal{D} =1$. In our case, this ratio is approximately $1.08$. This little deviation from unity is needed to keep $\theta_{13}$ nonzero and to maintain the non-maximality of $\theta_{23}$. A similar symmetry breaking operation of present kind with twisted parameters is motivated in certain extra dimension inspired theories \citep{Haba:2006dz}.  
\end{itemize}

In order to keep the discussion simple, all the parameters are treated as real and we hope that a further motivation in the same line is obtainable from the works in the Refs.\,\citep{Fritzsch:2006sm,Morisi:2011pm, Dev:2011qy,Adhikary:2013bma}. 

To summarize, we have put forward an ansatz to relate the neutrino mixing angles and the mass ratios in the similar way it happens for quarks and discussed the phenomenological consequences.  Finally we have associated the ansatz with certain sum rules to obtain possible patterns of neutrino mass matrix with exact texture, in the basis the charged lepton mass matrix is diagonal.

\begin{table*}
\setlength{\tabcolsep}{1 em}
\begin{center}
\begin{tabular}{l|c|c|c}
\hline 
\hline\\
 & {\footnotesize \textbf{Parameters} }& {\footnotesize \textbf{Prediction}} & {\footnotesize \textbf{Texture}} \\ 
\hline 
\hline\\
{\footnotesize \textbf{NH-ND}} & \begin{tabular}{l}
$\epsilon = 0.151859$,\\ 
$f=3.53882$, \\ 
$d=4.96783$, \\  
$s=1.21652$ ,\\ 
$m_{3}=0.0483055\,eV$. \\ 
\end{tabular}  & \begin{tabular}{l}
$m_{1}=0.00135518\,eV$,\\ 
$m_{2}=0.00892393\,eV$, \\ 
$\Delta\,m_{21}^2= 7.78\times 10^{-5}\,eV^{2}$, \\ 
$\Delta\,m_{31}^2=2.33\times 10^{-3}\,eV^{2}$,\\ 
$\sin^2\theta_{12}=0.289$, \\ 
$\sin^2\theta_{23}= 0.569$, \\ 
$\sin^2\theta_{13}=0.0233$, \\ 
$\Sigma\,m_{i}=0.0585846 \,eV$. \\ 
\end{tabular}  & 
\begin{tabular}{l}
$\begin {bmatrix}\mathcal {A} & 2 \,\mathcal {A} - \mathcal {C} & \
\mathcal {C} \\ 2\,\mathcal {A} - \mathcal {C} & 6\,\mathcal {A} + \mathcal {C} & 4\, \
\mathcal {A} + \mathcal {C} \\ \mathcal {C} & 4\, \mathcal {A} + \
\mathcal {C} & 5\, \mathcal {A} + \mathcal {C} \\
\end {bmatrix}m_{3}$\\ 
\\
\begin{scriptsize}
$\mathcal{A}=0.0946855$,\,\, $\mathcal{C}=0.038355$.
\end{scriptsize}
\end{tabular}
\\
\\
{\footnotesize \textbf{NH-QD}} & \begin{tabular}{l}
\begin{tabular}{l}
$\eta=0.762386$ ,\\ 
$b=1.56568$,\\ 
$c=0.96663$,\\ 
$m_{3}=0.0588\,eV$. \\ 
\end{tabular}  \\ 
\\
\\
\begin{tabular}{l}
$\eta= 0.756875$, \\ 
$b=1.4945$,\\ 
$c=0.96663$, \\ 
$m_{3}=0.0588\,eV$. \\ 
\end{tabular}  \\ 
\end{tabular}   & \begin{tabular}{l}
\begin{tabular}{l}
$m_{1}=0.0330417\,eV$,\\ 
$m_{2}=0.0341791\,eV$, \\ 
$\Delta\,m_{21}^2= 7.65\times 10^{-5}\,eV^{2}$, \\ 
$\Delta\,m_{31}^2=2.37\times 10^{-3}\,eV^{2}$,\\ 
$\sin^2\theta_{12}=0.315$, \\ 
$\sin^2\theta_{23}= 0.581$, \\ 
$\sin^2\theta_{13}=0.0319$, \\ 
$\Sigma\,m_{i}=0.126036 \,eV$. \\ 
\end{tabular} \\ 
\\
\\
\begin{tabular}{l}
$m_{1}=0.0325638\,eV$,\\ 
$m_{2}=0.0336859\,eV$, \\ 
$\Delta\,m_{21}^2= 7.43\times 10^{-5}\,eV^{2}$, \\ 
$\Delta\,m_{31}^2=2.40\times 10^{-3}\,eV^{2}$,\\ 
$\sin^2\theta_{12}=0.314$, \\ 
$\sin^2\theta_{23}= 0.573$, \\ 
$\sin^2\theta_{13}=0.0259$, \\ 
$\Sigma\,m_{i}=0.12506 \,eV$. \\ 
\end{tabular}  \\ 
\end{tabular}  & \begin{tabular}{l}
$\begin {bmatrix}
\mathcal {A} & \mathcal {B} &\frac{2}{3}\mathcal{B}\\
 \mathcal {B} & 4 \mathcal {E} & \mathcal {E} \\
\frac{2}{3}\mathcal{B} & \mathcal {E} & 4 \mathcal {E} - \mathcal {B} \\
\end{bmatrix}m_{3}$
\\
\\ 
\begin{scriptsize}
$\mathcal{A}=0.581819$,\,\, $\mathcal{B}=0.0636482$,\end{scriptsize}\\
\begin{scriptsize}
$\mathcal{E}=0.203162$.
\end{scriptsize}
\\
\\
\\
$\begin {bmatrix} 
\mathcal {A} & \mathcal {B} &\frac {2 } {3} \mathcal{B} \\ 
\mathcal {B} & \mathcal {D} &\mathcal {E} \\
\frac {2 }{3}\mathcal {B} &\mathcal {E} & \mathcal {D} - \mathcal {B} \\
\end{bmatrix}m_{3}$\\ 
\\ 
\begin{scriptsize}
$\mathcal{A}=0.571196 $,\,\, $\mathcal{B}=0.058653 $,\end{scriptsize}\\
\begin{scriptsize}
$\mathcal{D}=0.80716$,\,\, $ \mathcal{E}=0.208885$. 
\end{scriptsize}
\end{tabular} \\
\hline 
\\
{\footnotesize \textbf{IH-QD}} & \begin{tabular}{l}
\begin{tabular}{l}
$\eta=0.774459$,\\ 
$b=0.950921$,\\ 
$c=0.989598$, \\ 
$m_{2}=0.0612\,eV$. \\ 
\end{tabular}  \\ 
\\
\\
\begin{tabular}{l}
$\eta= 0.761187$, \\ 
$b=1.31518$,\\ 
$c=0.989553$, \\ 
$m_{3}=0.06047\,eV$. \\ 
\end{tabular}  \\ 
\end{tabular} & \begin{tabular}{l}
\begin{tabular}{l}
$m_{1}=0.0605614\,eV$,\\ 
$m_{3}=0.0367071\,eV$, \\ 
$\Delta\,m_{21}^2= 7.75\times 10^{-5}\,eV^{2}$, \\ 
$\Delta\,m_{31}^2=2.32\times 10^{-3}\,eV^{2}$,\\ 
$\sin^2\theta_{12}=0.3264$, \\ 
$\sin^2\theta_{23}= 0.599$, \\ 
$\sin^2\theta_{13}=0.0151$, \\ 
$\Sigma\,m_{i}=0.158464 \,eV$. \\ 
\end{tabular}\\ 
\\
\\
\begin{tabular}{l}
$m_{1}=0.0598426\,eV$,\\ 
$m_{3}=0.0350405\,eV$, \\ 
$\Delta\,m_{21}^2= 7.57\times 10^{-5}\,eV^{2}$, \\ 
$\Delta\,m_{31}^2=2.35\times 10^{-3}\,eV^{2}$,\\ 
$\sin^2\theta_{12}=0.3264$, \\ 
$\sin^2\theta_{23}= 0.579$, \\ 
$\sin^2\theta_{13}=0.0219$, \\ 
$\Sigma\,m_{i}=0.155355\,eV$. \\ 
\end{tabular}  \\ 
\end{tabular}   & \begin{tabular}{l}
$\begin {bmatrix} 
\mathcal {A} &\mathcal {B} & \mathcal {B} \\ 
\mathcal {B} &\mathcal { 
      D} &\mathcal {D} - \mathcal {A} - \mathcal {B} \\ 
B &\mathcal {D} - \mathcal {A} - \mathcal {B} &  \mathcal {F} \\
\end {bmatrix}m_{2}$ \\
\\ 
\begin{scriptsize}
$\mathcal{A}= 0.987095,\,\,\mathcal{B}=-0.0341297$
\end{scriptsize}\\
\begin{scriptsize}
$\mathcal{D}=0.761603 ,\,\mathcal{F}=0.840778$.
\end{scriptsize} 
\\
\\
$\begin {bmatrix} 
\mathcal {A} &\mathcal {B} & \mathcal {B} \\ 
\mathcal {B} &\mathcal { 
      D} &\mathcal {F} - \mathcal {A} + \mathcal {B} \\ 
B &\mathcal {F} - \mathcal {A} + \mathcal {B} &  \mathcal {F} \\
\end {bmatrix}m_{2}$\\
\\ 
\begin{scriptsize}
$\mathcal{A}= 0.983997,\,\,\mathcal{B}=-0.0430051$
\end{scriptsize}\\
\begin{scriptsize}
$\mathcal{D}=0.759459 ,\,\mathcal{F}=0.825692$,
\end{scriptsize} 
\end{tabular}  \\ 
\hline 
\end{tabular}
\caption{\label{table4}\footnotesize The summary of parametrization and the neutrino mass matrix texture}
\end{center} 
\end{table*}
\begin{table*}
\setlength{\tabcolsep}{1 em}
\begin{center}
\begin{tabular}{l||l}
\hline
\hline\\
\vspace{0.1 cm}
\textbf{I} & \begin{tabular}{l}$
\mathcal{A}\underbrace{\begin{bmatrix}
0 & 2 & 0\\
2 & 6 & 4\\
0 & 4 & 5
\end{bmatrix}}+
\mathcal{C}\left\lbrace\underbrace{\begin{bmatrix}
0 & 0 & 1\\
0 & 1 & 0\\
1 & 0 & 0
\end{bmatrix}} + \underbrace{\begin{bmatrix}
1 & 0 & 0\\
0 & 0 & 1\\
0 & 1 & 0
\end{bmatrix}}+ \underbrace{\begin{bmatrix}
0 & -1 & 0\\
-1 & 0 & 0\\
0 & 0 & 1
\end{bmatrix}}\right\rbrace+(\mathcal{A-C})\underbrace{\begin{bmatrix}
1 & 0 & 0\\
0 & 0 & 0\\
0 & 0 & 0
\end{bmatrix}}$\\
$\hspace{0.83 cm}H\hspace{1.9cm} f_{1} \hspace{1.455 cm} d_{1}\hspace{1.63cm} e_{4}\hspace{3.15 cm} T_{1}$
\end{tabular}\\
\hline\\
\textbf{II} &\begin{tabular}{l}$
(\mathcal{E-B})\underbrace{\begin{bmatrix}
1 & 0 & 0\\
0 & 0 & 1\\
0 & 1 & 0
\end{bmatrix}}+
(4\mathcal{E}-B)\underbrace{\begin{bmatrix}
1 & 0 & 0\\
0 & 1 & 0\\
0 & 0 & 1
\end{bmatrix}} + \mathcal{B}\underbrace{\begin{bmatrix}
1 & 1 & 1\\
1 & 1 & 1\\
1 & 1 & 1
\end{bmatrix}}+(\mathcal{A}-5\mathcal{E}+\frac{10}{9}\mathcal{B})\underbrace{\begin{bmatrix}
1 & 0 & 0\\
0 & 0 & 0\\
0 & 0 & 0
\end{bmatrix}} + \mathcal{B}\underbrace{\begin{bmatrix}
-\frac{1}{9} & 0 & -\frac{1}{3}\\
0 & 0 & 0\\
-\frac{1}{3} & 0& -1 
\end{bmatrix}}$\\
$\hspace{1.7 cm}d_{1} \hspace{1.6cm}  \hspace{1.07cm} I \hspace{1.7 cm} D\hspace{3.71cm} T_{1}\hspace{1.87 cm} T_{2}$
\end{tabular}\\
\hline\\
\textbf{III} &\begin{tabular}{l}$
(\mathcal{E-B})\underbrace{\begin{bmatrix}
1 & 0 & 0\\
0 & 0 & 1\\
0 & 1 & 0
\end{bmatrix}}+
(\mathcal{D}-B)\underbrace{\begin{bmatrix}
1 & 0 & 0\\
0 & 1 & 0\\
0 & 0 & 1
\end{bmatrix}} + \mathcal{B}\underbrace{\begin{bmatrix}
1 & 1 & 1\\
1 & 1 & 1\\
1 & 1 & 1
\end{bmatrix}}+(\mathcal{A}-\mathcal{E-D}+\frac{10}{9}\mathcal{B})\underbrace{\begin{bmatrix}
1 & 0 & 0\\
0 & 0 & 0\\
0 & 0 & 0
\end{bmatrix}} + \mathcal{B}\underbrace{\begin{bmatrix}
-\frac{1}{9} & 0 & -\frac{1}{3}\\
0 & 0 & 0\\
-\frac{1}{3} & 0& -1 
\end{bmatrix}}$\\
$\hspace{1.7 cm}d_{1} \hspace{1.6cm}  \hspace{0.99 cm} I \hspace{1.73 cm} D\hspace{4.15 cm} T_{1}\hspace{1.9 cm}T_{2}$
\end{tabular}\\
\hline\\
\textbf{IV} &\begin{tabular}{l}$
(\mathcal{A-D})\underbrace{\begin{bmatrix}
1 & 0 & 0\\
0 & 0 & -1\\
0 & -1 & 0
\end{bmatrix}}
- \mathcal{B}\left\lbrace\underbrace{\begin{bmatrix}
0 & -1 & 0\\
0 & 0 & 1\\
-1 & 0 & 0
\end{bmatrix}}+\underbrace{\begin{bmatrix}
0 & 0 & -1\\
-1 & 0 & 0\\
0 & 1 & 0
\end{bmatrix}}\right\rbrace+\mathcal{D}\underbrace{\begin{bmatrix}
1 & 0 & 0\\
0 & 1 & 0\\
0 & 0 & \frac{\mathcal{F}}{\mathcal{D}}
\end{bmatrix}}$\\
$\hspace{2.05 cm}d_{2} \hspace{2.4 cm} c_{3} \hspace{2 cm} b_{3}\hspace{2.29 cm} T_{3}$
\end{tabular}\\
\hline\\
\textbf{V} &\begin{tabular}{l}$
(\mathcal{A-F})\underbrace{\begin{bmatrix}
1 & 0 & 0\\
0 & 0 & -1\\
0 & -1 & 0
\end{bmatrix}}
+ \mathcal{B}\left\lbrace\underbrace{\begin{bmatrix}
1 & 1 & 1\\
1 & 1 & 1\\
1 & 1 & 1
\end{bmatrix}}-\underbrace{\begin{bmatrix}
1 & 0 & 0\\
0 & 1 & 0\\
0 & 0 & 1
\end{bmatrix}}\right\rbrace+\mathcal{F}\underbrace{\begin{bmatrix}
1 & 0 & 0\\
0 & \frac{\mathcal{D}}{\mathcal{F}} & 0\\
0 & 0 & 1
\end{bmatrix}}$\\
$\hspace{2.04cm}d_{2} \hspace{2.12 cm}D \hspace{1.48 cm} I \hspace{2.25 cm} T_{4}$
\end{tabular}\\
\\
\hline
\end{tabular} 
\caption{\label{table5} The expression of \textbf{Textures I-V}, in terms of corresponding building block matrices}
\end{center}
\end{table*}

\bibliography{mybib}

\begin{thebibliography}{0}%
\makeatletter
\providecommand \@ifxundefined [1]{%
 \@ifx{#1\undefined}
}%
\providecommand \@ifnum [1]{%
 \ifnum #1\expandafter \@firstoftwo
 \else \expandafter \@secondoftwo
 \fi
}%
\providecommand \@ifx [1]{%
 \ifx #1\expandafter \@firstoftwo
 \else \expandafter \@secondoftwo
 \fi
}%
\providecommand \natexlab [1]{#1}%
\providecommand \enquote  [1]{``#1''}%
\providecommand \bibnamefont  [1]{#1}%
\providecommand \bibfnamefont [1]{#1}%
\providecommand \citenamefont [1]{#1}%
\providecommand \href@noop [0]{\@secondoftwo}%
\providecommand \href [0]{\begingroup \@sanitize@url \@href}%
\providecommand \@href[1]{\@@startlink{#1}\@@href}%
\providecommand \@@href[1]{\endgroup#1\@@endlink}%
\providecommand \@sanitize@url [0]{\catcode `\\12\catcode `\$12\catcode
  `\&12\catcode `\#12\catcode `\^12\catcode `\_12\catcode `\%12\relax}%
\providecommand \@@startlink[1]{}%
\providecommand \@@endlink[0]{}%
\providecommand \url  [0]{\begingroup\@sanitize@url \@url }%
\providecommand \@url [1]{\endgroup\@href {#1}{\urlprefix }}%
\providecommand \urlprefix  [0]{URL }%
\providecommand \Eprint [0]{\href }%
\providecommand \doibase [0]{http://dx.doi.org/}%
\providecommand \selectlanguage [0]{\@gobble}%
\providecommand \bibinfo  [0]{\@secondoftwo}%
\providecommand \bibfield  [0]{\@secondoftwo}%
\providecommand \translation [1]{[#1]}%
\providecommand \BibitemOpen [0]{}%
\providecommand \bibitemStop [0]{}%
\providecommand \bibitemNoStop [0]{.\EOS\space}%
\providecommand \EOS [0]{\spacefactor3000\relax}%
\providecommand \BibitemShut  [1]{\csname bibitem#1\endcsname}%
\let\auto@bib@innerbib\@empty
\end{thebibliography}%


\begin{thebibliography}{11}%
\makeatletter
\providecommand \@ifxundefined [1]{%
 \@ifx{#1\undefined}
}%
\providecommand \@ifnum [1]{%
 \ifnum #1\expandafter \@firstoftwo
 \else \expandafter \@secondoftwo
 \fi
}%
\providecommand \@ifx [1]{%
 \ifx #1\expandafter \@firstoftwo
 \else \expandafter \@secondoftwo
 \fi
}%
\providecommand \natexlab [1]{#1}%
\providecommand \enquote  [1]{``#1''}%
\providecommand \bibnamefont  [1]{#1}%
\providecommand \bibfnamefont [1]{#1}%
\providecommand \citenamefont [1]{#1}%
\providecommand \href@noop [0]{\@secondoftwo}%
\providecommand \href [0]{\begingroup \@sanitize@url \@href}%
\providecommand \@href[1]{\@@startlink{#1}\@@href}%
\providecommand \@@href[1]{\endgroup#1\@@endlink}%
\providecommand \@sanitize@url [0]{\catcode `\\12\catcode `\$12\catcode
  `\&12\catcode `\#12\catcode `\^12\catcode `\_12\catcode `\%12\relax}%
\providecommand \@@startlink[1]{}%
\providecommand \@@endlink[0]{}%
\providecommand \url  [0]{\begingroup\@sanitize@url \@url }%
\providecommand \@url [1]{\endgroup\@href {#1}{\urlprefix }}%
\providecommand \urlprefix  [0]{URL }%
\providecommand \Eprint [0]{\href }%
\providecommand \doibase [0]{http://dx.doi.org/}%
\providecommand \selectlanguage [0]{\@gobble}%
\providecommand \bibinfo  [0]{\@secondoftwo}%
\providecommand \bibfield  [0]{\@secondoftwo}%
\providecommand \translation [1]{[#1]}%
\providecommand \BibitemOpen [0]{}%
\providecommand \bibitemStop [0]{}%
\providecommand \bibitemNoStop [0]{.\EOS\space}%
\providecommand \EOS [0]{\spacefactor3000\relax}%
\providecommand \BibitemShut  [1]{\csname bibitem#1\endcsname}%
\let\auto@bib@innerbib\@empty
\bibitem [{\citenamefont {Maki}\ \emph {et~al.}(1962)\citenamefont {Maki},
  \citenamefont {Nakagawa},\ and\ \citenamefont {Sakata}}]{Maki:1962mu}%
  \BibitemOpen
  \bibfield  {author} {\bibinfo {author} {\bibfnamefont {Z.}~\bibnamefont
  {Maki}}, \bibinfo {author} {\bibfnamefont {M.}~\bibnamefont {Nakagawa}}, \
  and\ \bibinfo {author} {\bibfnamefont {S.}~\bibnamefont {Sakata}},\ }\href
  {\doibase 10.1143/PTP.28.870} {\bibfield  {journal} {\bibinfo  {journal}
  {Prog. Theor. Phys.}\ }\textbf {\bibinfo {volume} {28}},\ \bibinfo {pages}
  {870} (\bibinfo {year} {1962})}\BibitemShut {NoStop}%
\bibitem [{\citenamefont {Gonzalez-Garcia}\ \emph {et~al.}(2014)\citenamefont
  {Gonzalez-Garcia}, \citenamefont {Maltoni},\ and\ \citenamefont
  {Schwetz}}]{Gonzalez-Garcia:2014bfa}%
  \BibitemOpen
  \bibfield  {author} {\bibinfo {author} {\bibfnamefont {M.~C.}\ \bibnamefont
  {Gonzalez-Garcia}}, \bibinfo {author} {\bibfnamefont {M.}~\bibnamefont
  {Maltoni}}, \ and\ \bibinfo {author} {\bibfnamefont {T.}~\bibnamefont
  {Schwetz}},\ }\href {\doibase 10.1007/JHEP11(2014)052} {\bibfield  {journal}
  {\bibinfo  {journal} {JHEP}\ }\textbf {\bibinfo {volume} {11}},\ \bibinfo
  {pages} {052} (\bibinfo {year} {2014})},\ \Eprint
  {http://arxiv.org/abs/1409.5439} {arXiv:1409.5439 [hep-ph]} \BibitemShut
  {NoStop}%
\bibitem [{\citenamefont {Forero}\ \emph {et~al.}(2014)\citenamefont {Forero},
  \citenamefont {Tortola},\ and\ \citenamefont {Valle}}]{Forero:2014bxa}%
  \BibitemOpen
  \bibfield  {author} {\bibinfo {author} {\bibfnamefont {D.~V.}\ \bibnamefont
  {Forero}}, \bibinfo {author} {\bibfnamefont {M.}~\bibnamefont {Tortola}}, \
  and\ \bibinfo {author} {\bibfnamefont {J.~W.~F.}\ \bibnamefont {Valle}},\
  }\bibfield  {booktitle} {\emph {\bibinfo {booktitle} {{Physical Review D 90
  (2014) 093006}}},\ }\href {\doibase 10.1103/PhysRevD.90.093006} {\bibfield
  {journal} {\bibinfo  {journal} {Phys. Rev.}\ }\textbf {\bibinfo {volume}
  {D90}},\ \bibinfo {pages} {093006} (\bibinfo {year} {2014})},\ \Eprint
  {http://arxiv.org/abs/1405.7540} {arXiv:1405.7540 [hep-ph]} \BibitemShut
  {NoStop}%
\bibitem [{\citenamefont {Harrison}\ \emph {et~al.}(2002)\citenamefont
  {Harrison}, \citenamefont {Perkins},\ and\ \citenamefont
  {Scott}}]{Harrison:2002er}%
  \BibitemOpen
  \bibfield  {author} {\bibinfo {author} {\bibfnamefont {P.~F.}\ \bibnamefont
  {Harrison}}, \bibinfo {author} {\bibfnamefont {D.~H.}\ \bibnamefont
  {Perkins}}, \ and\ \bibinfo {author} {\bibfnamefont {W.~G.}\ \bibnamefont
  {Scott}},\ }\href {\doibase 10.1016/S0370-2693(02)01336-9} {\bibfield
  {journal} {\bibinfo  {journal} {Phys. Lett.}\ }\textbf {\bibinfo {volume}
  {B530}},\ \bibinfo {pages} {167} (\bibinfo {year} {2002})},\ \Eprint
  {http://arxiv.org/abs/hep-ph/0202074} {arXiv:hep-ph/0202074 [hep-ph]}
  \BibitemShut {NoStop}%
\bibitem [{\citenamefont {Roy}\ and\ \citenamefont
  {Singh}(2015)}]{Roy:2015cza}%
  \BibitemOpen
  \bibfield  {author} {\bibinfo {author} {\bibfnamefont {S.}~\bibnamefont
  {Roy}}\ and\ \bibinfo {author} {\bibfnamefont {N.}~\bibnamefont {Singh}},\
  }\href {\doibase 10.1103/PhysRevD.91.096003} {\bibfield  {journal} {\bibinfo
  {journal} {Phys. Rev.}\ }\textbf {\bibinfo {volume} {D91}},\ \bibinfo {pages}
  {096003} (\bibinfo {year} {2015})}\BibitemShut {NoStop}%
\bibitem [{\citenamefont {Antusch}\ \emph {et~al.}(2014)\citenamefont
  {Antusch}, \citenamefont {King},\ and\ \citenamefont
  {Spinrath}}]{Antusch:2013rxa}%
  \BibitemOpen
  \bibfield  {author} {\bibinfo {author} {\bibfnamefont {S.}~\bibnamefont
  {Antusch}}, \bibinfo {author} {\bibfnamefont {S.~F.}\ \bibnamefont {King}}, \
  and\ \bibinfo {author} {\bibfnamefont {M.}~\bibnamefont {Spinrath}},\ }\href
  {\doibase 10.1103/PhysRevD.89.055027} {\bibfield  {journal} {\bibinfo
  {journal} {Phys. Rev.}\ }\textbf {\bibinfo {volume} {D89}},\ \bibinfo {pages}
  {055027} (\bibinfo {year} {2014})},\ \Eprint {http://arxiv.org/abs/1311.0877}
  {arXiv:1311.0877 [hep-ph]} \BibitemShut {NoStop}%
\bibitem [{\citenamefont {Kajiyama}\ \emph {et~al.}(2007)\citenamefont
  {Kajiyama}, \citenamefont {Raidal},\ and\ \citenamefont
  {Strumia}}]{Kajiyama:2007gx}%
  \BibitemOpen
  \bibfield  {author} {\bibinfo {author} {\bibfnamefont {Y.}~\bibnamefont
  {Kajiyama}}, \bibinfo {author} {\bibfnamefont {M.}~\bibnamefont {Raidal}}, \
  and\ \bibinfo {author} {\bibfnamefont {A.}~\bibnamefont {Strumia}},\ }\href
  {\doibase 10.1103/PhysRevD.76.117301} {\bibfield  {journal} {\bibinfo
  {journal} {Phys. Rev.}\ }\textbf {\bibinfo {volume} {D76}},\ \bibinfo {pages}
  {117301} (\bibinfo {year} {2007})},\ \Eprint {http://arxiv.org/abs/0705.4559}
  {arXiv:0705.4559 [hep-ph]} \BibitemShut {NoStop}%
\bibitem [{\citenamefont {Adulpravitchai}\ \emph {et~al.}(2009)\citenamefont
  {Adulpravitchai}, \citenamefont {Blum},\ and\ \citenamefont
  {Rodejohann}}]{Adulpravitchai:2009bg}%
  \BibitemOpen
  \bibfield  {author} {\bibinfo {author} {\bibfnamefont {A.}~\bibnamefont
  {Adulpravitchai}}, \bibinfo {author} {\bibfnamefont {A.}~\bibnamefont
  {Blum}}, \ and\ \bibinfo {author} {\bibfnamefont {W.}~\bibnamefont
  {Rodejohann}},\ }\href {\doibase 10.1088/1367-2630/11/6/063026} {\bibfield
  {journal} {\bibinfo  {journal} {New J. Phys.}\ }\textbf {\bibinfo {volume}
  {11}},\ \bibinfo {pages} {063026} (\bibinfo {year} {2009})},\ \Eprint
  {http://arxiv.org/abs/0903.0531} {arXiv:0903.0531 [hep-ph]} \BibitemShut
  {NoStop}%
\bibitem [{\citenamefont {Harrison}\ \emph {et~al.}(1995)\citenamefont
  {Harrison}, \citenamefont {Perkins},\ and\ \citenamefont
  {Scott}}]{Harrison:1994iv}%
  \BibitemOpen
  \bibfield  {author} {\bibinfo {author} {\bibfnamefont {P.~F.}\ \bibnamefont
  {Harrison}}, \bibinfo {author} {\bibfnamefont {D.~H.}\ \bibnamefont
  {Perkins}}, \ and\ \bibinfo {author} {\bibfnamefont {W.~G.}\ \bibnamefont
  {Scott}},\ }\href {\doibase 10.1016/0370-2693(95)00213-5} {\bibfield
  {journal} {\bibinfo  {journal} {Phys. Lett.}\ }\textbf {\bibinfo {volume}
  {B349}},\ \bibinfo {pages} {137} (\bibinfo {year} {1995})}\BibitemShut
  {NoStop}%
\bibitem [{\citenamefont {Meurant}(1974)}]{meurant1974noneuclidean}%
  \BibitemOpen
  \bibfield  {author} {\bibinfo {author} {\bibfnamefont {G.}~\bibnamefont
  {Meurant}},\ }\href {https://books.google.co.in/books?id=iLkzandfCc8C} {\emph
  {\bibinfo {title} {Noneuclidean tesselations and their groups}}},\ Pure and
  Applied Mathematics\ (\bibinfo  {publisher} {Elsevier Science},\ \bibinfo
  {year} {1974})\BibitemShut {NoStop}%
\bibitem [{\citenamefont {Lam}(2001)}]{Lam:2001fb}%
  \BibitemOpen
  \bibfield  {author} {\bibinfo {author} {\bibfnamefont {C.~S.}\ \bibnamefont
  {Lam}},\ }\href {\doibase 10.1016/S0370-2693(01)00465-8} {\bibfield
  {journal} {\bibinfo  {journal} {Phys. Lett.}\ }\textbf {\bibinfo {volume}
  {B507}},\ \bibinfo {pages} {214} (\bibinfo {year} {2001})},\ \Eprint
  {http://arxiv.org/abs/hep-ph/0104116} {arXiv:hep-ph/0104116 [hep-ph]}
  \BibitemShut {NoStop}%
\end{thebibliography}%


\begin{thebibliography}{44}%
\makeatletter
\providecommand \@ifxundefined [1]{%
 \@ifx{#1\undefined}
}%
\providecommand \@ifnum [1]{%
 \ifnum #1\expandafter \@firstoftwo
 \else \expandafter \@secondoftwo
 \fi
}%
\providecommand \@ifx [1]{%
 \ifx #1\expandafter \@firstoftwo
 \else \expandafter \@secondoftwo
 \fi
}%
\providecommand \natexlab [1]{#1}%
\providecommand \enquote  [1]{``#1''}%
\providecommand \bibnamefont  [1]{#1}%
\providecommand \bibfnamefont [1]{#1}%
\providecommand \citenamefont [1]{#1}%
\providecommand \href@noop [0]{\@secondoftwo}%
\providecommand \href [0]{\begingroup \@sanitize@url \@href}%
\providecommand \@href[1]{\@@startlink{#1}\@@href}%
\providecommand \@@href[1]{\endgroup#1\@@endlink}%
\providecommand \@sanitize@url [0]{\catcode `\\12\catcode `\$12\catcode
  `\&12\catcode `\#12\catcode `\^12\catcode `\_12\catcode `\%12\relax}%
\providecommand \@@startlink[1]{}%
\providecommand \@@endlink[0]{}%
\providecommand \url  [0]{\begingroup\@sanitize@url \@url }%
\providecommand \@url [1]{\endgroup\@href {#1}{\urlprefix }}%
\providecommand \urlprefix  [0]{URL }%
\providecommand \Eprint [0]{\href }%
\providecommand \doibase [0]{http://dx.doi.org/}%
\providecommand \selectlanguage [0]{\@gobble}%
\providecommand \bibinfo  [0]{\@secondoftwo}%
\providecommand \bibfield  [0]{\@secondoftwo}%
\providecommand \translation [1]{[#1]}%
\providecommand \BibitemOpen [0]{}%
\providecommand \bibitemStop [0]{}%
\providecommand \bibitemNoStop [0]{.\EOS\space}%
\providecommand \EOS [0]{\spacefactor3000\relax}%
\providecommand \BibitemShut  [1]{\csname bibitem#1\endcsname}%
\let\auto@bib@innerbib\@empty
\bibitem [{\citenamefont {Fukuyama}\ and\ \citenamefont
  {Nishiura}(1997)}]{Fukuyama:1997ky}%
  \BibitemOpen
  \bibfield  {author} {\bibinfo {author} {\bibfnamefont {T.}~\bibnamefont
  {Fukuyama}}\ and\ \bibinfo {author} {\bibfnamefont {H.}~\bibnamefont
  {Nishiura}},\ }\href@noop {} {\  (\bibinfo {year} {1997})},\ \Eprint
  {http://arxiv.org/abs/hep-ph/9702253} {arXiv:hep-ph/9702253 [hep-ph]}
  \BibitemShut {NoStop}%
\bibitem [{\citenamefont {Mohapatra}\ and\ \citenamefont
  {Nussinov}(1999)}]{Mohapatra:1998ka}%
  \BibitemOpen
  \bibfield  {author} {\bibinfo {author} {\bibfnamefont {R.~N.}\ \bibnamefont
  {Mohapatra}}\ and\ \bibinfo {author} {\bibfnamefont {S.}~\bibnamefont
  {Nussinov}},\ }\href {\doibase 10.1103/PhysRevD.60.013002} {\bibfield
  {journal} {\bibinfo  {journal} {Phys.Rev.}\ }\textbf {\bibinfo {volume}
  {D60}},\ \bibinfo {pages} {013002} (\bibinfo {year} {1999})},\ \Eprint
  {http://arxiv.org/abs/hep-ph/9809415} {arXiv:hep-ph/9809415 [hep-ph]}
  \BibitemShut {NoStop}%
\bibitem [{\citenamefont {Lam}(2001)}]{Lam:2001fb}%
  \BibitemOpen
  \bibfield  {author} {\bibinfo {author} {\bibfnamefont {C.}~\bibnamefont
  {Lam}},\ }\href {\doibase 10.1016/S0370-2693(01)00465-8} {\bibfield
  {journal} {\bibinfo  {journal} {Phys.Lett.}\ }\textbf {\bibinfo {volume}
  {B507}},\ \bibinfo {pages} {214} (\bibinfo {year} {2001})},\ \Eprint
  {http://arxiv.org/abs/hep-ph/0104116} {arXiv:hep-ph/0104116 [hep-ph]}
  \BibitemShut {NoStop}%
\bibitem [{\citenamefont {Mohapatra}\ \emph {et~al.}(2006)\citenamefont
  {Mohapatra}, \citenamefont {Nasri},\ and\ \citenamefont
  {Yu}}]{Mohapatra:2006pu}%
  \BibitemOpen
  \bibfield  {author} {\bibinfo {author} {\bibfnamefont {R.}~\bibnamefont
  {Mohapatra}}, \bibinfo {author} {\bibfnamefont {S.}~\bibnamefont {Nasri}}, \
  and\ \bibinfo {author} {\bibfnamefont {H.-B.}\ \bibnamefont {Yu}},\ }\href
  {\doibase 10.1016/j.physletb.2006.06.032} {\bibfield  {journal} {\bibinfo
  {journal} {Phys.Lett.}\ }\textbf {\bibinfo {volume} {B639}},\ \bibinfo
  {pages} {318} (\bibinfo {year} {2006})},\ \Eprint
  {http://arxiv.org/abs/hep-ph/0605020} {arXiv:hep-ph/0605020 [hep-ph]}
  \BibitemShut {NoStop}%
\bibitem [{\citenamefont {Harrison}\ and\ \citenamefont
  {Scott}(2002)}]{Harrison:2002et}%
  \BibitemOpen
  \bibfield  {author} {\bibinfo {author} {\bibfnamefont {P.}~\bibnamefont
  {Harrison}}\ and\ \bibinfo {author} {\bibfnamefont {W.}~\bibnamefont
  {Scott}},\ }\href {\doibase 10.1016/S0370-2693(02)02772-7} {\bibfield
  {journal} {\bibinfo  {journal} {Phys.Lett.}\ }\textbf {\bibinfo {volume}
  {B547}},\ \bibinfo {pages} {219} (\bibinfo {year} {2002})},\ \Eprint
  {http://arxiv.org/abs/hep-ph/0210197} {arXiv:hep-ph/0210197 [hep-ph]}
  \BibitemShut {NoStop}%
\bibitem [{\citenamefont {Grimus}\ and\ \citenamefont
  {Lavoura}(2001)}]{Grimus:2001ex}%
  \BibitemOpen
  \bibfield  {author} {\bibinfo {author} {\bibfnamefont {W.}~\bibnamefont
  {Grimus}}\ and\ \bibinfo {author} {\bibfnamefont {L.}~\bibnamefont
  {Lavoura}},\ }\href {\doibase 10.1088/1126-6708/2001/07/045} {\bibfield
  {journal} {\bibinfo  {journal} {JHEP}\ }\textbf {\bibinfo {volume} {0107}},\
  \bibinfo {pages} {045} (\bibinfo {year} {2001})},\ \Eprint
  {http://arxiv.org/abs/hep-ph/0105212} {arXiv:hep-ph/0105212 [hep-ph]}
  \BibitemShut {NoStop}%
\bibitem [{\citenamefont {Grimus}\ and\ \citenamefont
  {Lavoura}(2003)}]{Grimus:2003kq}%
  \BibitemOpen
  \bibfield  {author} {\bibinfo {author} {\bibfnamefont {W.}~\bibnamefont
  {Grimus}}\ and\ \bibinfo {author} {\bibfnamefont {L.}~\bibnamefont
  {Lavoura}},\ }\href {\doibase 10.1016/j.physletb.2003.08.032} {\bibfield
  {journal} {\bibinfo  {journal} {Phys.Lett.}\ }\textbf {\bibinfo {volume}
  {B572}},\ \bibinfo {pages} {189} (\bibinfo {year} {2003})},\ \Eprint
  {http://arxiv.org/abs/hep-ph/0305046} {arXiv:hep-ph/0305046 [hep-ph]}
  \BibitemShut {NoStop}%
\bibitem [{\citenamefont {Kitabayashi}\ and\ \citenamefont
  {Yasue}(2003)}]{Kitabayashi:2002jd}%
  \BibitemOpen
  \bibfield  {author} {\bibinfo {author} {\bibfnamefont {T.}~\bibnamefont
  {Kitabayashi}}\ and\ \bibinfo {author} {\bibfnamefont {M.}~\bibnamefont
  {Yasue}},\ }\href {\doibase 10.1103/PhysRevD.67.015006} {\bibfield  {journal}
  {\bibinfo  {journal} {Phys.Rev.}\ }\textbf {\bibinfo {volume} {D67}},\
  \bibinfo {pages} {015006} (\bibinfo {year} {2003})},\ \Eprint
  {http://arxiv.org/abs/hep-ph/0209294} {arXiv:hep-ph/0209294 [hep-ph]}
  \BibitemShut {NoStop}%
\bibitem [{\citenamefont {Koide}(2004)}]{Koide:2003rx}%
  \BibitemOpen
  \bibfield  {author} {\bibinfo {author} {\bibfnamefont {Y.}~\bibnamefont
  {Koide}},\ }\href {\doibase 10.1103/PhysRevD.69.093001} {\bibfield  {journal}
  {\bibinfo  {journal} {Phys.Rev.}\ }\textbf {\bibinfo {volume} {D69}},\
  \bibinfo {pages} {093001} (\bibinfo {year} {2004})},\ \Eprint
  {http://arxiv.org/abs/hep-ph/0312207} {arXiv:hep-ph/0312207 [hep-ph]}
  \BibitemShut {NoStop}%
\bibitem [{\citenamefont {Ma}\ and\ \citenamefont
  {Rajasekaran}(2001)}]{Ma:2001dn}%
  \BibitemOpen
  \bibfield  {author} {\bibinfo {author} {\bibfnamefont {E.}~\bibnamefont
  {Ma}}\ and\ \bibinfo {author} {\bibfnamefont {G.}~\bibnamefont
  {Rajasekaran}},\ }\href {\doibase 10.1103/PhysRevD.64.113012} {\bibfield
  {journal} {\bibinfo  {journal} {Phys.Rev.}\ }\textbf {\bibinfo {volume}
  {D64}},\ \bibinfo {pages} {113012} (\bibinfo {year} {2001})},\ \Eprint
  {http://arxiv.org/abs/hep-ph/0106291} {arXiv:hep-ph/0106291 [hep-ph]}
  \BibitemShut {NoStop}%
\bibitem [{\citenamefont {Ma}(2002)}]{Ma:2002yp}%
  \BibitemOpen
  \bibfield  {author} {\bibinfo {author} {\bibfnamefont {E.}~\bibnamefont
  {Ma}},\ }\href {\doibase 10.1142/S0217732302006722} {\bibfield  {journal}
  {\bibinfo  {journal} {Mod.Phys.Lett.}\ }\textbf {\bibinfo {volume} {A17}},\
  \bibinfo {pages} {627} (\bibinfo {year} {2002})},\ \Eprint
  {http://arxiv.org/abs/hep-ph/0203238} {arXiv:hep-ph/0203238 [hep-ph]}
  \BibitemShut {NoStop}%
\bibitem [{\citenamefont {Ma}(2004)}]{Ma:2004zv}%
  \BibitemOpen
  \bibfield  {author} {\bibinfo {author} {\bibfnamefont {E.}~\bibnamefont
  {Ma}},\ }\href {\doibase 10.1103/PhysRevD.70.031901} {\bibfield  {journal}
  {\bibinfo  {journal} {Phys.Rev.}\ }\textbf {\bibinfo {volume} {D70}},\
  \bibinfo {pages} {031901} (\bibinfo {year} {2004})},\ \Eprint
  {http://arxiv.org/abs/hep-ph/0404199} {arXiv:hep-ph/0404199 [hep-ph]}
  \BibitemShut {NoStop}%
\bibitem [{\citenamefont {Ma}(2006)}]{Ma:2005qf}%
  \BibitemOpen
  \bibfield  {author} {\bibinfo {author} {\bibfnamefont {E.}~\bibnamefont
  {Ma}},\ }\href {\doibase 10.1103/PhysRevD.73.057304} {\bibfield  {journal}
  {\bibinfo  {journal} {Phys.Rev.}\ }\textbf {\bibinfo {volume} {D73}},\
  \bibinfo {pages} {057304} (\bibinfo {year} {2006})},\ \Eprint
  {http://arxiv.org/abs/hep-ph/0511133} {arXiv:hep-ph/0511133 [hep-ph]}
  \BibitemShut {NoStop}%
\bibitem [{\citenamefont {Altarelli}\ and\ \citenamefont
  {Feruglio}(2005)}]{Altarelli:2005yp}%
  \BibitemOpen
  \bibfield  {author} {\bibinfo {author} {\bibfnamefont {G.}~\bibnamefont
  {Altarelli}}\ and\ \bibinfo {author} {\bibfnamefont {F.}~\bibnamefont
  {Feruglio}},\ }\href {\doibase 10.1016/j.nuclphysb.2005.05.005} {\bibfield
  {journal} {\bibinfo  {journal} {Nucl.Phys.}\ }\textbf {\bibinfo {volume}
  {B720}},\ \bibinfo {pages} {64} (\bibinfo {year} {2005})},\ \Eprint
  {http://arxiv.org/abs/hep-ph/0504165} {arXiv:hep-ph/0504165 [hep-ph]}
  \BibitemShut {NoStop}%
\bibitem [{\citenamefont {Altarelli}\ and\ \citenamefont
  {Feruglio}(2006)}]{Altarelli:2005yx}%
  \BibitemOpen
  \bibfield  {author} {\bibinfo {author} {\bibfnamefont {G.}~\bibnamefont
  {Altarelli}}\ and\ \bibinfo {author} {\bibfnamefont {F.}~\bibnamefont
  {Feruglio}},\ }\href {\doibase 10.1016/j.nuclphysb.2006.02.015} {\bibfield
  {journal} {\bibinfo  {journal} {Nucl.Phys.}\ }\textbf {\bibinfo {volume}
  {B741}},\ \bibinfo {pages} {215} (\bibinfo {year} {2006})},\ \Eprint
  {http://arxiv.org/abs/hep-ph/0512103} {arXiv:hep-ph/0512103 [hep-ph]}
  \BibitemShut {NoStop}%
\bibitem [{\citenamefont {Grimus}\ \emph {et~al.}(2009)\citenamefont {Grimus},
  \citenamefont {Lavoura},\ and\ \citenamefont {Ludl}}]{Grimus:2009pg}%
  \BibitemOpen
  \bibfield  {author} {\bibinfo {author} {\bibfnamefont {W.}~\bibnamefont
  {Grimus}}, \bibinfo {author} {\bibfnamefont {L.}~\bibnamefont {Lavoura}}, \
  and\ \bibinfo {author} {\bibfnamefont {P.}~\bibnamefont {Ludl}},\ }\href
  {\doibase 10.1088/0954-3899/36/11/115007} {\bibfield  {journal} {\bibinfo
  {journal} {J.Phys.}\ }\textbf {\bibinfo {volume} {G36}},\ \bibinfo {pages}
  {115007} (\bibinfo {year} {2009})},\ \Eprint {http://arxiv.org/abs/0906.2689}
  {arXiv:0906.2689 [hep-ph]} \BibitemShut {NoStop}%
\bibitem [{\citenamefont {Zee}(2005)}]{Zee:2005ut}%
  \BibitemOpen
  \bibfield  {author} {\bibinfo {author} {\bibfnamefont {A.}~\bibnamefont
  {Zee}},\ }\href {\doibase 10.1016/j.physletb.2005.09.068} {\bibfield
  {journal} {\bibinfo  {journal} {Phys.Lett.}\ }\textbf {\bibinfo {volume}
  {B630}},\ \bibinfo {pages} {58} (\bibinfo {year} {2005})},\ \Eprint
  {http://arxiv.org/abs/hep-ph/0508278} {arXiv:hep-ph/0508278 [hep-ph]}
  \BibitemShut {NoStop}%
\bibitem [{\citenamefont {Adhikary}\ and\ \citenamefont
  {Ghosal}(2007)}]{Adhikary:2006jx}%
  \BibitemOpen
  \bibfield  {author} {\bibinfo {author} {\bibfnamefont {B.}~\bibnamefont
  {Adhikary}}\ and\ \bibinfo {author} {\bibfnamefont {A.}~\bibnamefont
  {Ghosal}},\ }\href {\doibase 10.1103/PhysRevD.75.073020} {\bibfield
  {journal} {\bibinfo  {journal} {Phys.Rev.}\ }\textbf {\bibinfo {volume}
  {D75}},\ \bibinfo {pages} {073020} (\bibinfo {year} {2007})},\ \Eprint
  {http://arxiv.org/abs/hep-ph/0609193} {arXiv:hep-ph/0609193 [hep-ph]}
  \BibitemShut {NoStop}%
\bibitem [{\citenamefont {Ahn}\ \emph {et~al.}(2012)\citenamefont {Ahn} \emph
  {et~al.}}]{Ahn:2012nd}%
  \BibitemOpen
  \bibfield  {author} {\bibinfo {author} {\bibfnamefont {J.}~\bibnamefont
  {Ahn}} \emph {et~al.} (\bibinfo {collaboration} {RENO collaboration}),\
  }\href {\doibase 10.1103/PhysRevLett.108.191802} {\bibfield  {journal}
  {\bibinfo  {journal} {Phys.Rev.Lett.}\ }\textbf {\bibinfo {volume} {108}},\
  \bibinfo {pages} {191802} (\bibinfo {year} {2012})},\ \Eprint
  {http://arxiv.org/abs/1204.0626} {arXiv:1204.0626 [hep-ex]} \BibitemShut
  {NoStop}%
\bibitem [{\citenamefont {Abe}\ \emph {et~al.}(2012)\citenamefont {Abe} \emph
  {et~al.}}]{Abe:2011fz}%
  \BibitemOpen
  \bibfield  {author} {\bibinfo {author} {\bibfnamefont {Y.}~\bibnamefont
  {Abe}} \emph {et~al.} (\bibinfo {collaboration} {DOUBLE-CHOOZ
  Collaboration}),\ }\href {\doibase 10.1103/PhysRevLett.108.131801} {\bibfield
   {journal} {\bibinfo  {journal} {Phys.Rev.Lett.}\ }\textbf {\bibinfo {volume}
  {108}},\ \bibinfo {pages} {131801} (\bibinfo {year} {2012})},\ \Eprint
  {http://arxiv.org/abs/1112.6353} {arXiv:1112.6353 [hep-ex]} \BibitemShut
  {NoStop}%
\bibitem [{\citenamefont {An}\ \emph {et~al.}(2012)\citenamefont {An} \emph
  {et~al.}}]{An:2012eh}%
  \BibitemOpen
  \bibfield  {author} {\bibinfo {author} {\bibfnamefont {F.}~\bibnamefont {An}}
  \emph {et~al.} (\bibinfo {collaboration} {DAYA-BAY Collaboration}),\ }\href
  {\doibase 10.1103/PhysRevLett.108.171803} {\bibfield  {journal} {\bibinfo
  {journal} {Phys.Rev.Lett.}\ }\textbf {\bibinfo {volume} {108}},\ \bibinfo
  {pages} {171803} (\bibinfo {year} {2012})},\ \Eprint
  {http://arxiv.org/abs/1203.1669} {arXiv:1203.1669 [hep-ex]} \BibitemShut
  {NoStop}%
\bibitem [{\citenamefont {Forero}\ \emph {et~al.}(2014)\citenamefont {Forero},
  \citenamefont {Tortola},\ and\ \citenamefont {Valle}}]{Forero:2014bxa}%
  \BibitemOpen
  \bibfield  {author} {\bibinfo {author} {\bibfnamefont {D.}~\bibnamefont
  {Forero}}, \bibinfo {author} {\bibfnamefont {M.}~\bibnamefont {Tortola}}, \
  and\ \bibinfo {author} {\bibfnamefont {J.}~\bibnamefont {Valle}},\
  }\href@noop {} {\  (\bibinfo {year} {2014})},\ \Eprint
  {http://arxiv.org/abs/1405.7540} {arXiv:1405.7540 [hep-ph]} \BibitemShut
  {NoStop}%
\bibitem [{\citenamefont {Adhikary}\ \emph
  {et~al.}(2013{\natexlab{a}})\citenamefont {Adhikary}, \citenamefont
  {Ghosal},\ and\ \citenamefont {Roy}}]{Adhikary:2012mt}%
  \BibitemOpen
  \bibfield  {author} {\bibinfo {author} {\bibfnamefont {B.}~\bibnamefont
  {Adhikary}}, \bibinfo {author} {\bibfnamefont {A.}~\bibnamefont {Ghosal}}, \
  and\ \bibinfo {author} {\bibfnamefont {P.}~\bibnamefont {Roy}},\ }\href
  {\doibase 10.1142/S0217751X13501182} {\bibfield  {journal} {\bibinfo
  {journal} {Int.J.Mod.Phys.}\ }\textbf {\bibinfo {volume} {A28}},\ \bibinfo
  {pages} {1350118} (\bibinfo {year} {2013}{\natexlab{a}})},\ \Eprint
  {http://arxiv.org/abs/1210.5328} {arXiv:1210.5328 [hep-ph]} \BibitemShut
  {NoStop}%
\bibitem [{\citenamefont {Grimus}\ and\ \citenamefont
  {Lavoura}(2013)}]{Grimus:2012hu}%
  \BibitemOpen
  \bibfield  {author} {\bibinfo {author} {\bibfnamefont {W.}~\bibnamefont
  {Grimus}}\ and\ \bibinfo {author} {\bibfnamefont {L.}~\bibnamefont
  {Lavoura}},\ }\href {\doibase 10.1002/prop.201200118} {\bibfield  {journal}
  {\bibinfo  {journal} {Fortsch.Phys.}\ }\textbf {\bibinfo {volume} {61}},\
  \bibinfo {pages} {535} (\bibinfo {year} {2013})},\ \Eprint
  {http://arxiv.org/abs/1207.1678} {arXiv:1207.1678 [hep-ph]} \BibitemShut
  {NoStop}%
\bibitem [{\citenamefont {Adhikary}\ \emph {et~al.}(2014)\citenamefont
  {Adhikary}, \citenamefont {Ghosal},\ and\ \citenamefont
  {Roy}}]{Adhikary:2013mfa}%
  \BibitemOpen
  \bibfield  {author} {\bibinfo {author} {\bibfnamefont {B.}~\bibnamefont
  {Adhikary}}, \bibinfo {author} {\bibfnamefont {A.}~\bibnamefont {Ghosal}}, \
  and\ \bibinfo {author} {\bibfnamefont {P.}~\bibnamefont {Roy}},\ }\href
  {\doibase 10.1007/s12648-014-0485-7} {\bibfield  {journal} {\bibinfo
  {journal} {Indian J.Phys.}\ }\textbf {\bibinfo {volume} {88}},\ \bibinfo
  {pages} {979} (\bibinfo {year} {2014})},\ \Eprint
  {http://arxiv.org/abs/1311.6746} {arXiv:1311.6746 [hep-ph]} \BibitemShut
  {NoStop}%
\bibitem [{\citenamefont {Cabibbo}(1978)}]{Cabibbo:1977nk}%
  \BibitemOpen
  \bibfield  {author} {\bibinfo {author} {\bibfnamefont {N.}~\bibnamefont
  {Cabibbo}},\ }\href {\doibase 10.1016/0370-2693(78)90132-6} {\bibfield
  {journal} {\bibinfo  {journal} {Phys.Lett.}\ }\textbf {\bibinfo {volume}
  {B72}},\ \bibinfo {pages} {333} (\bibinfo {year} {1978})}\BibitemShut
  {NoStop}%
\bibitem [{\citenamefont {Wolfenstein}(1983)}]{Wolfenstein:1983yz}%
  \BibitemOpen
  \bibfield  {author} {\bibinfo {author} {\bibfnamefont {L.}~\bibnamefont
  {Wolfenstein}},\ }\href {\doibase 10.1103/PhysRevLett.51.1945} {\bibfield
  {journal} {\bibinfo  {journal} {Phys.Rev.Lett.}\ }\textbf {\bibinfo {volume}
  {51}},\ \bibinfo {pages} {1945} (\bibinfo {year} {1983})}\BibitemShut
  {NoStop}%
\bibitem [{\citenamefont {Gatto}\ \emph {et~al.}(1968)\citenamefont {Gatto},
  \citenamefont {Sartori},\ and\ \citenamefont {Tonin}}]{Gatto:1968ss}%
  \BibitemOpen
  \bibfield  {author} {\bibinfo {author} {\bibfnamefont {R.}~\bibnamefont
  {Gatto}}, \bibinfo {author} {\bibfnamefont {G.}~\bibnamefont {Sartori}}, \
  and\ \bibinfo {author} {\bibfnamefont {M.}~\bibnamefont {Tonin}},\ }\href
  {\doibase 10.1016/0370-2693(68)90150-0} {\bibfield  {journal} {\bibinfo
  {journal} {Phys.Lett.}\ }\textbf {\bibinfo {volume} {B28}},\ \bibinfo {pages}
  {128} (\bibinfo {year} {1968})}\BibitemShut {NoStop}%
\bibitem [{\citenamefont {Fritzsch}(1999)}]{Fritzsch:1999yd}%
  \BibitemOpen
  \bibfield  {author} {\bibinfo {author} {\bibfnamefont {H.}~\bibnamefont
  {Fritzsch}},\ }\href@noop {} {\  (\bibinfo {year} {1999})},\ \Eprint
  {http://arxiv.org/abs/hep-ph/9901275} {arXiv:hep-ph/9901275 [hep-ph]}
  \BibitemShut {NoStop}%
\bibitem [{\citenamefont {Fritzsch}\ and\ \citenamefont
  {Xing}(2000)}]{Fritzsch:1999ee}%
  \BibitemOpen
  \bibfield  {author} {\bibinfo {author} {\bibfnamefont {H.}~\bibnamefont
  {Fritzsch}}\ and\ \bibinfo {author} {\bibfnamefont {Z.-z.}\ \bibnamefont
  {Xing}},\ }\href {\doibase 10.1016/S0146-6410(00)00102-2} {\bibfield
  {journal} {\bibinfo  {journal} {Prog.Part.Nucl.Phys.}\ }\textbf {\bibinfo
  {volume} {45}},\ \bibinfo {pages} {1} (\bibinfo {year} {2000})},\ \Eprint
  {http://arxiv.org/abs/hep-ph/9912358} {arXiv:hep-ph/9912358 [hep-ph]}
  \BibitemShut {NoStop}%
\bibitem [{\citenamefont {Harrison}\ \emph {et~al.}(2002)\citenamefont
  {Harrison}, \citenamefont {Perkins},\ and\ \citenamefont
  {Scott}}]{Harrison:2002er}%
  \BibitemOpen
  \bibfield  {author} {\bibinfo {author} {\bibfnamefont {P.}~\bibnamefont
  {Harrison}}, \bibinfo {author} {\bibfnamefont {D.}~\bibnamefont {Perkins}}, \
  and\ \bibinfo {author} {\bibfnamefont {W.}~\bibnamefont {Scott}},\ }\href
  {\doibase 10.1016/S0370-2693(02)01336-9} {\bibfield  {journal} {\bibinfo
  {journal} {Phys.Lett.}\ }\textbf {\bibinfo {volume} {B530}},\ \bibinfo
  {pages} {167} (\bibinfo {year} {2002})},\ \Eprint
  {http://arxiv.org/abs/hep-ph/0202074} {arXiv:hep-ph/0202074 [hep-ph]}
  \BibitemShut {NoStop}%
\bibitem [{\citenamefont {Harrison}\ and\ \citenamefont
  {Scott}(2004)}]{Harrison:2004uh}%
  \BibitemOpen
  \bibfield  {author} {\bibinfo {author} {\bibfnamefont {P.}~\bibnamefont
  {Harrison}}\ and\ \bibinfo {author} {\bibfnamefont {W.}~\bibnamefont
  {Scott}},\ }\href@noop {} {\ ,\ \bibinfo {pages} {435} (\bibinfo {year}
  {2004})},\ \Eprint {http://arxiv.org/abs/hep-ph/0402006}
  {arXiv:hep-ph/0402006 [hep-ph]} \BibitemShut {NoStop}%
\bibitem [{\citenamefont {Abbas}\ and\ \citenamefont
  {Smirnov}(2010)}]{Abbas:2010jw}%
  \BibitemOpen
  \bibfield  {author} {\bibinfo {author} {\bibfnamefont {M.}~\bibnamefont
  {Abbas}}\ and\ \bibinfo {author} {\bibfnamefont {A.~Y.}\ \bibnamefont
  {Smirnov}},\ }\href {\doibase 10.1103/PhysRevD.82.013008} {\bibfield
  {journal} {\bibinfo  {journal} {Phys.Rev.}\ }\textbf {\bibinfo {volume}
  {D82}},\ \bibinfo {pages} {013008} (\bibinfo {year} {2010})},\ \Eprint
  {http://arxiv.org/abs/1004.0099} {arXiv:1004.0099 [hep-ph]} \BibitemShut
  {NoStop}%
\bibitem [{\citenamefont {Fritzsch}(2012)}]{Fritzsch:2012zp}%
  \BibitemOpen
  \bibfield  {author} {\bibinfo {author} {\bibfnamefont {H.}~\bibnamefont
  {Fritzsch}},\ }\href {\doibase 10.1142/S0217732312500794} {\bibfield
  {journal} {\bibinfo  {journal} {Mod.Phys.Lett.}\ }\textbf {\bibinfo {volume}
  {A27}},\ \bibinfo {pages} {1250079} (\bibinfo {year} {2012})},\ \Eprint
  {http://arxiv.org/abs/1203.4460} {arXiv:1203.4460 [hep-ph]} \BibitemShut
  {NoStop}%
\bibitem [{\citenamefont {Meloni}\ \emph {et~al.}(2011)\citenamefont {Meloni},
  \citenamefont {Morisi},\ and\ \citenamefont {Peinado}}]{Meloni:2010aw}%
  \BibitemOpen
  \bibfield  {author} {\bibinfo {author} {\bibfnamefont {D.}~\bibnamefont
  {Meloni}}, \bibinfo {author} {\bibfnamefont {S.}~\bibnamefont {Morisi}}, \
  and\ \bibinfo {author} {\bibfnamefont {E.}~\bibnamefont {Peinado}},\ }\href
  {\doibase 10.1088/0954-3899/38/1/015003} {\bibfield  {journal} {\bibinfo
  {journal} {J.Phys.}\ }\textbf {\bibinfo {volume} {G38}},\ \bibinfo {pages}
  {015003} (\bibinfo {year} {2011})},\ \Eprint {http://arxiv.org/abs/1005.3482}
  {arXiv:1005.3482 [hep-ph]} \BibitemShut {NoStop}%
\bibitem [{\citenamefont {King}\ \emph {et~al.}(2014)\citenamefont {King},
  \citenamefont {Merle}, \citenamefont {Morisi}, \citenamefont {Shimizu},\ and\
  \citenamefont {Tanimoto}}]{King:2014nza}%
  \BibitemOpen
  \bibfield  {author} {\bibinfo {author} {\bibfnamefont {S.~F.}\ \bibnamefont
  {King}}, \bibinfo {author} {\bibfnamefont {A.}~\bibnamefont {Merle}},
  \bibinfo {author} {\bibfnamefont {S.}~\bibnamefont {Morisi}}, \bibinfo
  {author} {\bibfnamefont {Y.}~\bibnamefont {Shimizu}}, \ and\ \bibinfo
  {author} {\bibfnamefont {M.}~\bibnamefont {Tanimoto}},\ }\href {\doibase
  10.1088/1367-2630/16/4/045018} {\bibfield  {journal} {\bibinfo  {journal}
  {New J.Phys.}\ }\textbf {\bibinfo {volume} {16}},\ \bibinfo {pages} {045018}
  (\bibinfo {year} {2014})},\ \Eprint {http://arxiv.org/abs/1402.4271}
  {arXiv:1402.4271 [hep-ph]} \BibitemShut {NoStop}%
\bibitem [{\citenamefont {Ishimori}\ \emph {et~al.}(2011)\citenamefont
  {Ishimori}, \citenamefont {Shimizu}, \citenamefont {Tanimoto},\ and\
  \citenamefont {Watanabe}}]{Ishimori:2010fs}%
  \BibitemOpen
  \bibfield  {author} {\bibinfo {author} {\bibfnamefont {H.}~\bibnamefont
  {Ishimori}}, \bibinfo {author} {\bibfnamefont {Y.}~\bibnamefont {Shimizu}},
  \bibinfo {author} {\bibfnamefont {M.}~\bibnamefont {Tanimoto}}, \ and\
  \bibinfo {author} {\bibfnamefont {A.}~\bibnamefont {Watanabe}},\ }\href
  {\doibase 10.1103/PhysRevD.83.033004} {\bibfield  {journal} {\bibinfo
  {journal} {Phys.Rev.}\ }\textbf {\bibinfo {volume} {D83}},\ \bibinfo {pages}
  {033004} (\bibinfo {year} {2011})},\ \Eprint {http://arxiv.org/abs/1010.3805}
  {arXiv:1010.3805 [hep-ph]} \BibitemShut {NoStop}%
\bibitem [{\citenamefont {Shimizu}\ \emph {et~al.}(2011)\citenamefont
  {Shimizu}, \citenamefont {Tanimoto},\ and\ \citenamefont
  {Watanabe}}]{Shimizu:2011xg}%
  \BibitemOpen
  \bibfield  {author} {\bibinfo {author} {\bibfnamefont {Y.}~\bibnamefont
  {Shimizu}}, \bibinfo {author} {\bibfnamefont {M.}~\bibnamefont {Tanimoto}}, \
  and\ \bibinfo {author} {\bibfnamefont {A.}~\bibnamefont {Watanabe}},\ }\href
  {\doibase 10.1143/PTP.126.81} {\bibfield  {journal} {\bibinfo  {journal}
  {Prog.Theor.Phys.}\ }\textbf {\bibinfo {volume} {126}},\ \bibinfo {pages}
  {81} (\bibinfo {year} {2011})},\ \Eprint {http://arxiv.org/abs/1105.2929}
  {arXiv:1105.2929 [hep-ph]} \BibitemShut {NoStop}%
\bibitem [{\citenamefont {Adhikary}\ and\ \citenamefont
  {Roy}(2013)}]{Adhikary:2012zx}%
  \BibitemOpen
  \bibfield  {author} {\bibinfo {author} {\bibfnamefont {B.}~\bibnamefont
  {Adhikary}}\ and\ \bibinfo {author} {\bibfnamefont {P.}~\bibnamefont {Roy}},\
  }\href {\doibase 10.1155/2013/324756} {\bibfield  {journal} {\bibinfo
  {journal} {Adv.High Energy Phys.}\ }\textbf {\bibinfo {volume} {2013}},\
  \bibinfo {pages} {324756} (\bibinfo {year} {2013})},\ \Eprint
  {http://arxiv.org/abs/1211.0371} {arXiv:1211.0371 [hep-ph]} \BibitemShut
  {NoStop}%
\bibitem [{\citenamefont {Haba}\ \emph {et~al.}(2006)\citenamefont {Haba},
  \citenamefont {Watanabe},\ and\ \citenamefont {Yoshioka}}]{Haba:2006dz}%
  \BibitemOpen
  \bibfield  {author} {\bibinfo {author} {\bibfnamefont {N.}~\bibnamefont
  {Haba}}, \bibinfo {author} {\bibfnamefont {A.}~\bibnamefont {Watanabe}}, \
  and\ \bibinfo {author} {\bibfnamefont {K.}~\bibnamefont {Yoshioka}},\ }\href
  {\doibase 10.1103/PhysRevLett.97.041601} {\bibfield  {journal} {\bibinfo
  {journal} {Phys.Rev.Lett.}\ }\textbf {\bibinfo {volume} {97}},\ \bibinfo
  {pages} {041601} (\bibinfo {year} {2006})},\ \Eprint
  {http://arxiv.org/abs/hep-ph/0603116} {arXiv:hep-ph/0603116 [hep-ph]}
  \BibitemShut {NoStop}%
\bibitem [{\citenamefont {Fritzsch}\ and\ \citenamefont
  {Xing}(2006)}]{Fritzsch:2006sm}%
  \BibitemOpen
  \bibfield  {author} {\bibinfo {author} {\bibfnamefont {H.}~\bibnamefont
  {Fritzsch}}\ and\ \bibinfo {author} {\bibfnamefont {Z.-z.}\ \bibnamefont
  {Xing}},\ }\href {\doibase 10.1016/j.physletb.2006.02.028} {\bibfield
  {journal} {\bibinfo  {journal} {Phys.Lett.}\ }\textbf {\bibinfo {volume}
  {B634}},\ \bibinfo {pages} {514} (\bibinfo {year} {2006})},\ \Eprint
  {http://arxiv.org/abs/hep-ph/0601104} {arXiv:hep-ph/0601104 [hep-ph]}
  \BibitemShut {NoStop}%
\bibitem [{\citenamefont {Morisi}\ \emph {et~al.}(2011)\citenamefont {Morisi},
  \citenamefont {Patel},\ and\ \citenamefont {Peinado}}]{Morisi:2011pm}%
  \BibitemOpen
  \bibfield  {author} {\bibinfo {author} {\bibfnamefont {S.}~\bibnamefont
  {Morisi}}, \bibinfo {author} {\bibfnamefont {K.~M.}\ \bibnamefont {Patel}}, \
  and\ \bibinfo {author} {\bibfnamefont {E.}~\bibnamefont {Peinado}},\ }\href
  {\doibase 10.1103/PhysRevD.84.053002} {\bibfield  {journal} {\bibinfo
  {journal} {Phys.Rev.}\ }\textbf {\bibinfo {volume} {D84}},\ \bibinfo {pages}
  {053002} (\bibinfo {year} {2011})},\ \Eprint {http://arxiv.org/abs/1107.0696}
  {arXiv:1107.0696 [hep-ph]} \BibitemShut {NoStop}%
\bibitem [{\citenamefont {Dev}\ \emph {et~al.}(2011)\citenamefont {Dev},
  \citenamefont {Gupta},\ and\ \citenamefont {Gautam}}]{Dev:2011qy}%
  \BibitemOpen
  \bibfield  {author} {\bibinfo {author} {\bibfnamefont {S.}~\bibnamefont
  {Dev}}, \bibinfo {author} {\bibfnamefont {S.}~\bibnamefont {Gupta}}, \ and\
  \bibinfo {author} {\bibfnamefont {R.~R.}\ \bibnamefont {Gautam}},\ }\href
  {\doibase 10.1016/j.physletb.2011.06.055} {\bibfield  {journal} {\bibinfo
  {journal} {Phys.Lett.}\ }\textbf {\bibinfo {volume} {B702}},\ \bibinfo
  {pages} {28} (\bibinfo {year} {2011})},\ \Eprint
  {http://arxiv.org/abs/1106.3873} {arXiv:1106.3873 [hep-ph]} \BibitemShut
  {NoStop}%
\bibitem [{\citenamefont {Adhikary}\ \emph
  {et~al.}(2013{\natexlab{b}})\citenamefont {Adhikary}, \citenamefont
  {Chakraborty},\ and\ \citenamefont {Ghosal}}]{Adhikary:2013bma}%
  \BibitemOpen
  \bibfield  {author} {\bibinfo {author} {\bibfnamefont {B.}~\bibnamefont
  {Adhikary}}, \bibinfo {author} {\bibfnamefont {M.}~\bibnamefont
  {Chakraborty}}, \ and\ \bibinfo {author} {\bibfnamefont {A.}~\bibnamefont
  {Ghosal}},\ }\href {\doibase 10.1007/JHEP10(2013)043,
  10.1007/JHEP09(2014)180} {\bibfield  {journal} {\bibinfo  {journal} {JHEP}\
  }\textbf {\bibinfo {volume} {1310}},\ \bibinfo {pages} {043} (\bibinfo {year}
  {2013}{\natexlab{b}})},\ \Eprint {http://arxiv.org/abs/1307.0988}
  {arXiv:1307.0988 [hep-ph]} \BibitemShut {NoStop}%
\end{thebibliography}%

\end{document}